\documentclass[aps,prl,reprint,superscriptaddress,notitlepage]{revtex4-2}

\usepackage[dvipsnames]{xcolor}
\definecolor{myblue}{RGB}{41,  128, 128}
\definecolor{mywhit}{RGB}{237, 248, 177}
\definecolor{myorag}{RGB}{202, 86,  44}

\usepackage{chemformula} 
\usepackage{overpic}
\graphicspath{{Images/}}
\usepackage{subcaption}
\usepackage[outline]{contour}
\usepackage{xcolor}
\usepackage{mhchem}
\usepackage{siunitx}
\usepackage{miller}
\usepackage{gensymb}
\usepackage{float}
\usepackage{miller}
\usepackage{siunitx}

\usepackage{bm}
\newcommand{\uvec}[1]{\boldsymbol{\hat{\textbf{#1}}}}
\usepackage{soul}  
\sethlcolor{pink}
\begin{document}

\preprint{APS/123-QED}

\title{Experimental Evidence of a change of Exchange Anisotropy Sign with Temperature in Zn-Substituted \ce{Cu2OSeO3} }

\author{S. H. Moody}
 \email{samuel.h.moody@durham.ac.uk}
\address{Durham University, Department of Physics, South Road, Durham, DH1 3LE, United Kingdom}
\author{P. Nielsen}
\address{Durham University, Department of Physics, South Road, Durham, DH1 3LE, United Kingdom}
\author{M. N. Wilson}
\address{Durham University, Department of Physics, South Road, Durham, DH1 3LE, United Kingdom}
\author{D. Alba Venero}
 \affiliation{ISIS Neutron and Muon Source, Rutherford Appleton Laboratory, Didcot OX11 0QX, UK.}
\author{A. \v{S}tefan\v{c}i\v{c}}
\address{University of Warwick, Department of Physics, Coventry, CV4 7AL, United Kingdom}
\author{G. Balakrishnan}
\address{University of Warwick, Department of Physics, Coventry, CV4 7AL, United Kingdom}
\author{P. D. Hatton}
\address{Durham University, Department of Physics, South Road, Durham, DH1 3LE, United Kingdom}

\date{\today}

\begin{abstract}
We report small-angle neutron scattering from the conical state in a single crystal of Zn-substituted \ce{Cu2OSeO3}. Using a 3D vector-field magnet to reorient the conical wavevector, our measurements show that the magnitude of the conical wavevector changes as a function of crystallographic direction. These changes are caused by the anisotropic exchange interaction (AEI), whose magnitude transitions from a maxima to a minima along the \hkl<111> and \hkl<100> crystallographic directions respectively. We further find that the AEI constant undergoes a change of sign from positive to negative with decreasing temperature. Unlike in the related compound \ce{FeGe}, where similar behaviour of the AEI induces a reorientation of the helical wavevector, we show that the zero field helical wavevector in \ce{(Cu_{0.98}Zn_{0.02})_2OSeO3} remains along the \hkl<100> directions at all temperatures due to the competing fourth-order magnetocrystalline anisotropy becoming dominant at lower temperatures.

\end{abstract}

\maketitle


The lack of inversion symmetry in the family of B20 chiral cubic materials allows a non-vanishing Dzyaloshinskii-Moriya interaction (DMI), which competes with numerous other magnetic interactions to stabilize a wide variety of incommensurate magnetic states \cite{Lancaster2019, Dzya1964, Neubauer2009, Tanigaki2015, Wilhelm2011, Yu2011, Seki2012, Tokunaga2015}. At zero field and at temperatures below $T_\textnormal{C}$, the isotropic exchange interaction and DMI (which prefer spin-aligned with constant $A$ and spin-perpendicular with constant $D$ respectively), twist the magnetic texture into a helix with a periodicity determined by the ratio of the strengths of the two interactions, $q \approx D/A$. Typically, $q$ is close to 0.1~nm$^{-1}$, suitable for measurement with magnetic small angle neutron scattering (SANS) experiments. The non-centrosymmetric crystal structure allows for non-zero off-diagonal terms in the exchange tensor \cite{Blundell}, which gives rise to the so-called anisotropic exchange interaction (AEI) with a strength $\gamma$, where $|\gamma| < |A|$. In MnSi, the helical wavevector was shown to align along the \hkl<111> directions \cite{ISHIKAWA1976} due to a negative AEI constant \cite{Bak1980, Nakanishi1980, Maleyev2006, Ukleev2021}. $\gamma$ is also known to be temperature dependent, with a change of sign with temperature being responsible for the helical wavevector reorientation in the related material, FeGe \cite{Lebech_1989, Plumer_1990}.  \

The orientation of these incommensurate textures can also be manipulated by the application of a magnetic field, which induces a transition from the multi-domain magnetic helix into a single-domain magnetic cone state, whose wavevector typically follows the direction of the applied magnetic field \cite{ISHIKAWA1976}. As shown in Fig. 1a, in the cone state the spins cant towards the field direction to minimize the Zeeman interaction. Applying a higher field reduces the conical angle $\theta$, but maintains a constant wavevector, $q$,  until the cone angle reaches zero to form a forced ferromagnetic state \cite{ISHIKAWA1977}. \

The behaviour of chiral magnets adopting a B20 structure materials near $T_\text{C}$ is further enriched by thermal fluctuations, which induce a hexagonal lattice of magnetic skyrmions within small region at non-zero field. Magnetic skyrmions are vortex-like whirls of magnetization that are the 2D solitonic solutions to systems hosting a number of competing magnetic interactions \cite{ISHIMOTO1990, Muhlbauer2009, Lancaster2019, Robler2006, Robler2006, Vousden2016, Bogdanov2020}. Their particle-like nature and topological properties make magnetic skyrmions appealing for applications within a variety of spintronic devices, ranging from race-track memory schemes and logic devices, to stochastic and neuromorphic computing \cite{Zhang2015, Fert2017, Zhang2018, Zhang2015, Schaffer2019, Zhang2020, Li2017, Song2020}.
\begin{figure}
    \centering
    \includegraphics[width=0.6\linewidth, clip, trim={0 0cm 0 0}]{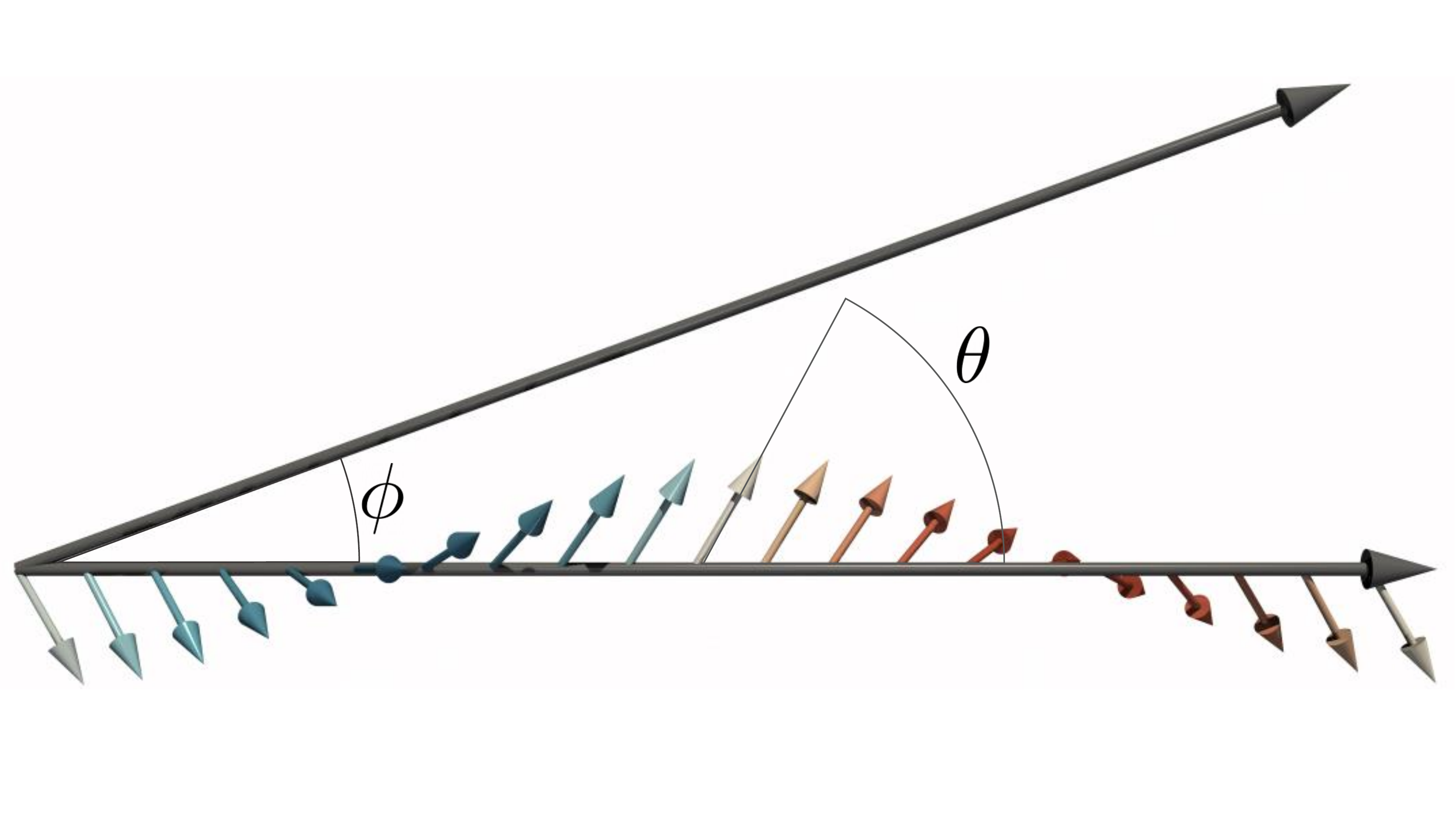}
    \includegraphics[width=0.98\linewidth,  clip, trim={0 0cm 0 0}]{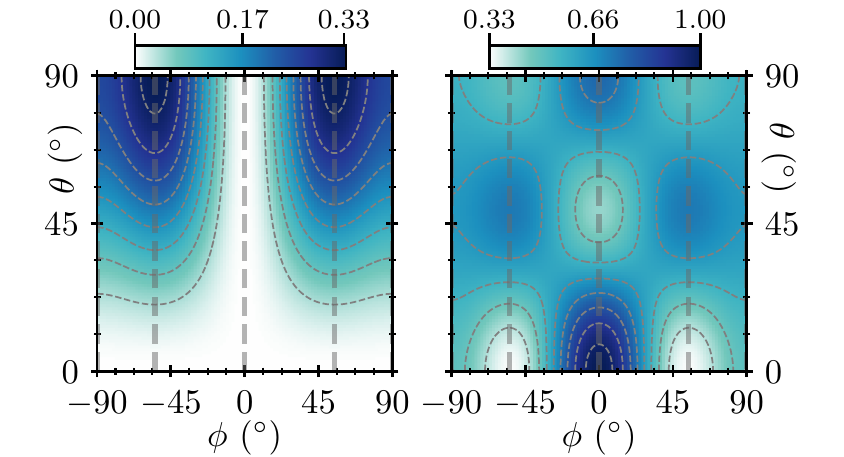}
    \begin{overpic}[width=0.75\linewidth,  clip, trim={0 3cm 0 0.5cm}]{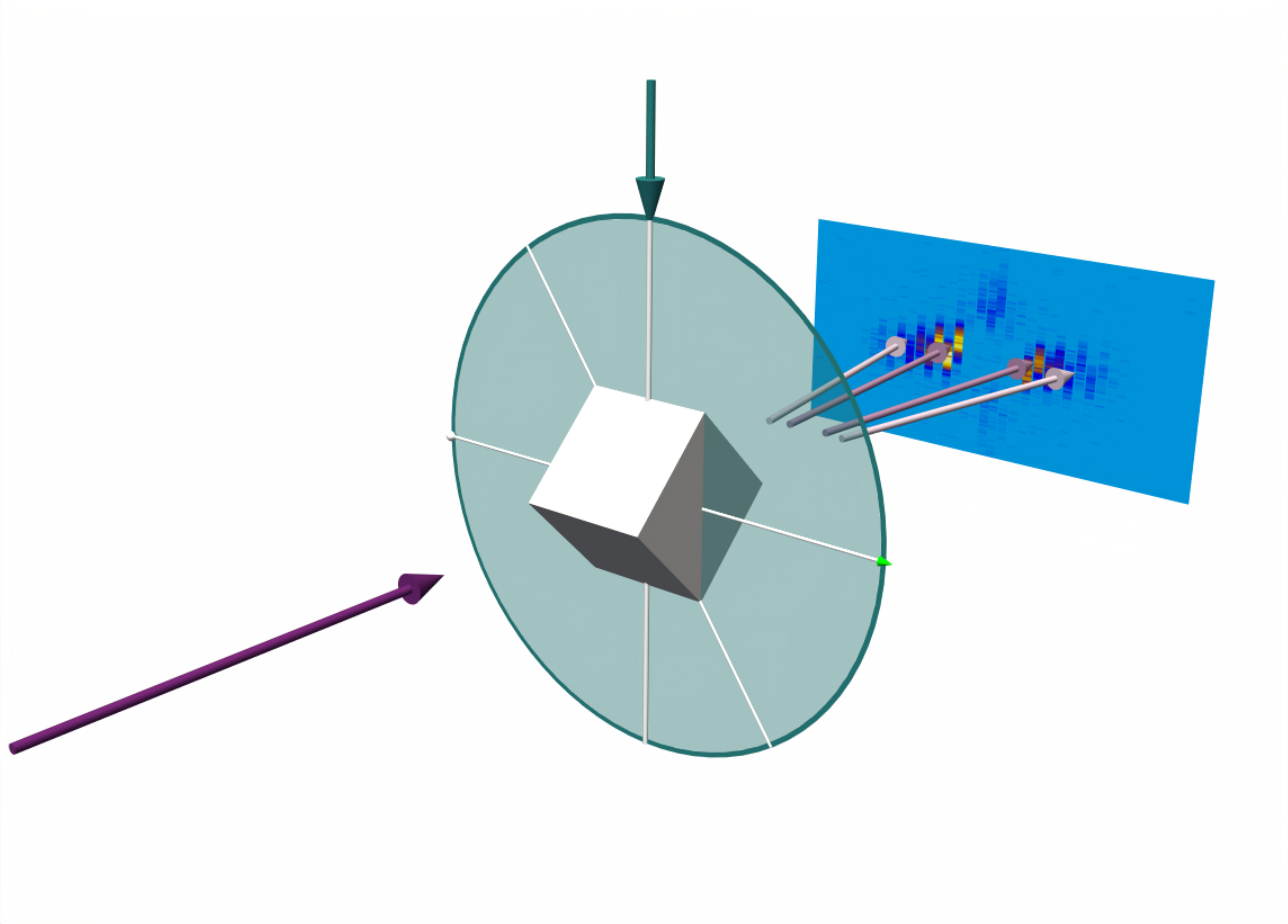}
    \put(-10, 168){\large\textbf{(a)}}
    \put(85.5, 174){\textbf{\hkl[110]}}
    \put(90, 147.5){\textbf{q}}
    \put(-10, 135){\large\textbf{(b)}}
    \put(0, 118){\color{white}{\textbf{(i)}}}
    \put(55, 118){\color{white}{\textbf{(ii)}}}
    \put(-10, 58){\large\textbf{(c)}}
    \put(18.5, 135){$E /\gamma $ }
    \put(70.5, 135){$E /K $ }
    \put(14, 15){$n_0$}
    \put(52, 55){$H$}
    \put(23, 30){\hkl[00-1]}    
    \put(30, 47){\hkl[1-11]}
	\put(51.5, 48.5){\hkl[110]}
    \end{overpic}
    \caption{\small \centering a) One period of the magnetic conical state with wavevector, \textbf{q}, where localized moments cant towards \textbf{q} by the angle $\theta$. The direction of \textbf{q} is given by the polar coordinate $\phi$, with $\phi=0, 55, 90^{\circ}$ being the \hkl[110], \hkl[11-1] and \hkl[00-1] directions respectively. b) AEI(i) and MCA(ii) energy landscapes of a conical state normalised by the constants $\gamma$ and $K$. c) SANS scattering geometry with 3D vector magnet. Incoming neutrons ($n_0$) parallel to the \hkl[1-10] direction, with the applied magnetic field perpendicular to $n_0$ within the scattering plane (Blue Circle). }
    \label{fig:1}
\end{figure}
\begin{figure*}
\begin{center}
    \centering
    \begin{overpic}[width=0.99\linewidth]{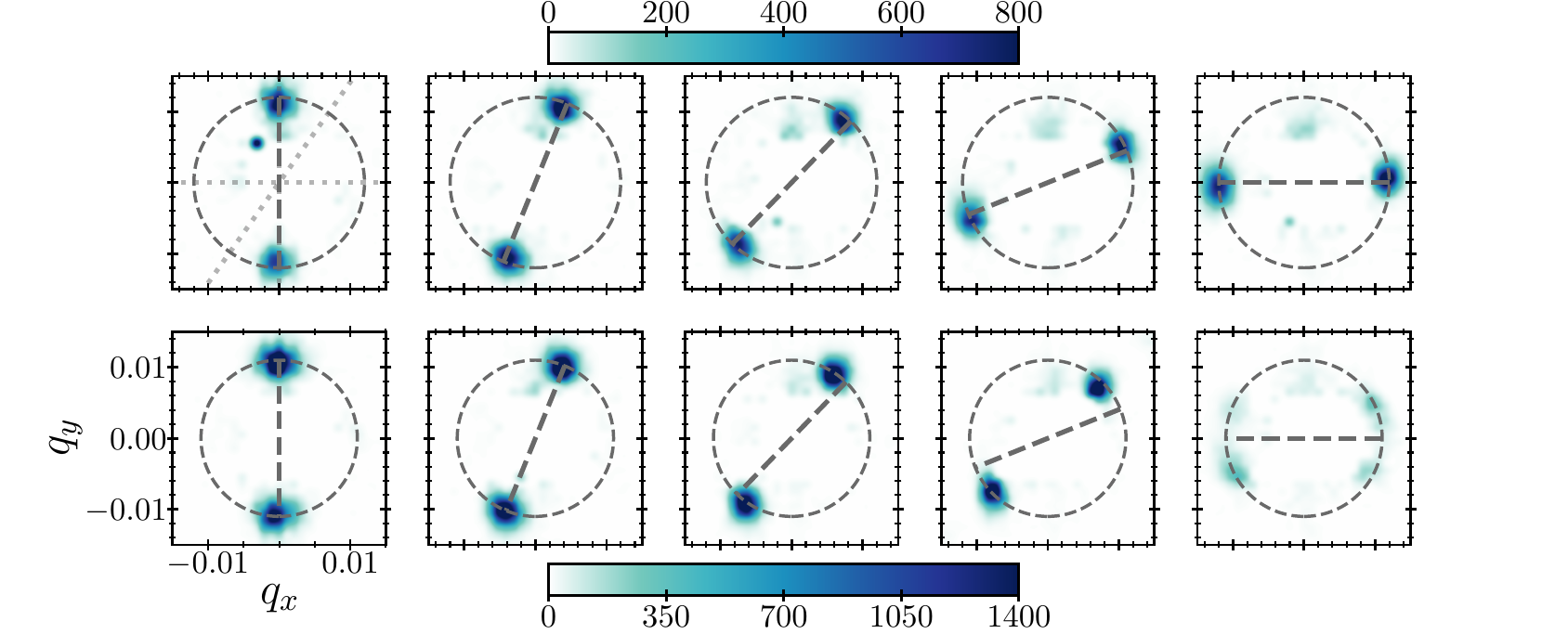}
    \put(7, 36){\large \textbf{(a)}}
    \put(7, 20){\large \textbf{(b)}}
    \put(48, 41){ {$I_{50\textnormal{K}}$ (a.u)}}
    \put(48, -1){{$I_{5\textnormal{K}}$ (a.u)}}
    \put(11.3, 22.5){50 K}
    \put(11.3, 6){5 K}
    \put(19.3, 22.5){40 mT}
    \put(19.3, 6){70 mT}
    \put(11.5, 34){(i)}
    \put(27.5, 34){(ii)}
    \put(44, 34){(iii)}
    \put(60, 34){(iv)}
    \put(76.5, 34){(v)}
    \put(11.5, 17.2){(i)}
    \put(27.5, 17.2){(ii)}
    \put(44, 17.2){(iii)}
    \put(60, 17.2){(iv)}
    \put(76.5, 17.2){(v)}
    \put(15.5, 36.5){\hkl[110]}
    \put(20.8, 36.5){\hkl[11-1]}
    \put(20.3, 29.1){\hkl[00-1]}
    \end{overpic}
    \caption{\small Selection of reduced SANS patterns for field rotations at 40 mT, 50 K(a) and 70 mT, 5 K(b). The field angles for each frame are 0, 22, 44, 68 and 90$^{\circ}$ for frames (i-v) and are shown by the thick dashed lines. The thin-dashed circle gives the location of equidistant $q=0.011$ $\si{\angstrom}^{-1}$. Crystallographic directions are shown by the dotted lines in a(i). }
    \label{fig:YZ Scans}
\end{center}
\end{figure*}

The effects of anisotropic interactions such as magnetocrystalline anisotropy (MCA) and the AEI on these magnetic textures are typically overlooked, as the inherent cubic symmetry of the B20 materials limits the strength of these energy terms \cite{Sucksmith1954}. Despite this, anisotropy-stabilised magnetic textures within the multiferroic B20 material \ce{Cu2OSeO3} have been found to exist at low temperatures, such as the low-temperature skyrmion (LTS) \cite{Chacon2018, Bannenberg2019, Halder2018, Aqeel2021} and tilted conical (TC) phase, where the conical wavevector deviates from the direction of the magnetic field applied along a \hkl[100], with $q$ canting towards the \hkl<111> \cite{Qian2018}. The forms of the MCA and AEI for a magnetic cone with angle $\theta$ and wavevector-angle $\phi$ are shown in Fig.~1b. In (ii), the local minima along the \hkl<111> directions for $K > 0$ explain the origin of the TC, but an additional AEI shown in (i) is essential to prevent the texture collapsing into a field-polarized state. Currently, the effects of the AEI have not been investigated on the LTS, and decoupling the effects of MCA and the AEI remains a challenge as the conventional method of torque magnetometry only investigates the field-polarized state where the AEI vanishes \cite{preissinger2020vital}. Distinguishing and quantifying these anisotropic interactions is therefore important to better understand the low-temperature behaviour of \ce{Cu2OSeO3}. \

In this letter, we utilize a 3D vector-field magnet together with small-angle neutron scattering (SANS) to show that the AEI changes sign from positive to negative with decreasing temperature in \ce{(Cu_{0.98}Zn_{0.02})_2OSeO3}. Unlike in the related compound FeGe, this change of sign does not induce a reorientation of the zero-field magnetic helix. This demonstrates that the 4$^\textnormal{th}$ order MCA also increases in magnitude with decreasing temperature, greater that the AEI in order to stabilize the zero-field magnetic helices along the \hkl<100> directions at all temperatures \cite{Birch2020}. Our findings provide insight into the low temperature behavior of \ce{Cu2OSeO3} that will be essential for complete understanding of the creation and stabilization mechanisms of the recently discovered LTS and TC phases within pristine \ce{Cu2OSeO3}. \

A 20.5 mg single crystal of \ce{(Cu_{0.98}Zn_{0.02})_2OSeO3} was grown at the University of Warwick utilizing the chemical vapour transport technique, see \cite{Moody2018} for details. For the SANS experiment, the LARMOR instrument at the ISIS Pulsed Neutron and Muon Source was used. The sample was aligned with an x-ray Laue camera (Multiwire Laboratories) such that the \hkl[110] direction was vertical in the laboratory frame and the \hkl[1 -1 0 ] direction was parallel with the incident neutron beam. This orientation allows the magnetic textures whose periodic components lie within a plane spanned by the all 3 high cubic-symmetry directions (\hkl<100>, \hkl<110>, and \hkl<111>) to simultaneously satisfy the Bragg condition \cite{MB2019}, see Fig.~1c. 
A zinc-substituted sample was chosen for our study due to the reduction of the exchange interaction compared to pristine \ce{Cu_2OSeO_3}, resulting in a greater value of $q$  \cite{Birch2020}. Previously, we have shown strong similarities of the magnetic behavior between Zn-substituted and pristine samples \cite{Birch2020}, and hence we expect our results here will extend to pristine samples. The time-of-flight neutron diffraction data was reduced with Mantid \cite{ARNOLD2014156}, using scattered neutron wavelengths between 0.9 and 13.5 \si{\angstrom}. \

To observe the effects of the anisotropic interactions on the magnetic textures within \ce{Cu2OSeO3}, we performed field scans by rotating the direction of the field at a constant magnitude. Initially, the field was applied vertically in the laboratory frame after zero-field cooling. The field orientation was then rotated 90 degrees in 46 steps about an axis parallel with the neutron beam, such that the \hkl[1 -1 0], \hkl[1 1 -1], \hkl[0 0 -1] directions were parallel with the field at angles of 0, 35.3 and 90 degrees respectively. This procedure was performed at 5, 12 and 50 K at magnetic field magnitudes of 70, 60 and 40 mT respectively. These fields were chosen to avoid phase coexistence between the magnetic conical and helical states, see Supplementary Material [URL will be inserted by publisher]. A selection of frames from the field scans are shown in Fig.~2a. \

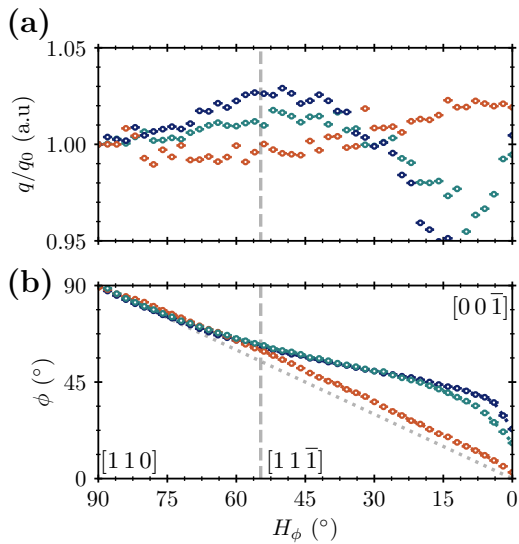
\begin{figure}[b]
    \centering
    \begin{overpic}[trim={0cm 0cm 0 0}, clip, width=0.85\linewidth]{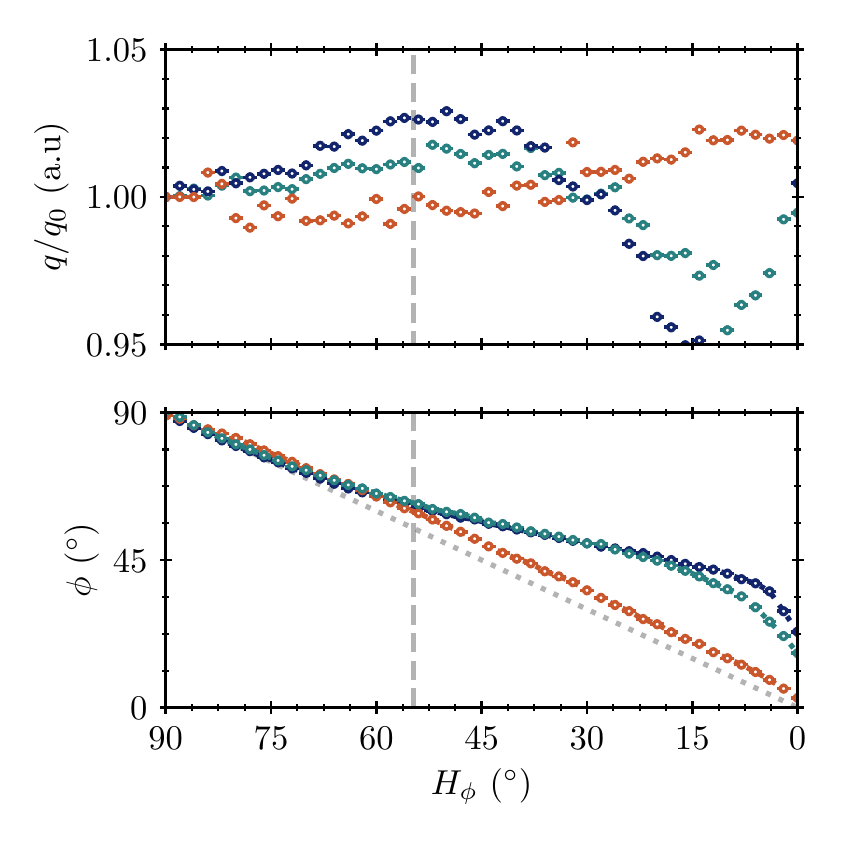}
    \put(3,97){\large\textbf{(a)}}
    \put(3,50){\large\textbf{(b)}}
    \put(20,18){\hkl[110]}
    \put(50,18){\hkl[11-1]}
    \put(83.5,46){\hkl[00-1]}
	\end{overpic}  
	\vspace{-12pt}
	\caption{\small a) Normalised wavevector magnitude, and wavevector angle (b), $\phi$ in Fig.~1a, of the conical state as a function of applied field angle, $H_\phi$.
	Dark blue, light blue and orange indicate temperatures of 5, 12, and 50 K respectively. Dashed line shows \hkl[11-1] direction. Dotted line shows $H_\phi$.}  
    \label{fig:2Dfits}
\end{figure}

In the first frame of the 50 K dataset, Fig.~2a(i), the vertically applied magnetic field induces a magnetic conical state with wavevector along the \hkl[1 -1 0] direction, which we detect as a single pair of vertical peaks with $q \approx 0.011 \si{\angstrom}^{-1}$, corresponding to a real space length of 57 nm. Upon rotating the field, Fig.~2a(ii-v), the conical wavevector rotates in an attempt to follow the direction of the magnetic field. This behaviour is expected in the limit in which the AEI and MCA are much smaller than the isotropic Dzyaloshinskii-Moriya and exchange interactions ($\gamma,K << A, D$), allowing the conical wavevector to rotate freely to minimize the Zeeman interaction. 
However, this behaviour is not replicated within the data taken at lower temperatures in Fig.~2b(i-v), where angle of the conical wavevector ($\phi$) lags behind the magnetic field angle ($H_\phi$), particularly after the field passes through the \hkl[11-1] direction. This offset between the field direction and $q$ increases in magnitude with further rotation of the magnetic field, up to a maxima just before the field is applied along the \hkl[00-1] direction. This magnetic state is characteristic of the tilted conical state seen previously in pristine \ce{Cu2OSeO3} \cite{Qian2018}, and shows that while isotropic interactions dominate at temperatures near $T_\textnormal{C}$, the anisotropic interactions become increasingly important at low temperatures and prevent free rotation of the magnetic conical state. \

In order to quantify the effects of the anisotropic interactions, the conical peaks collected during the field scans shown in Fig.~2 and an additional set at 12 K were fit using two dimensional Gaussian functions in polar coordinates, allowing the extraction of the magnetic wavevector angle, $\phi$, and magnitude, $q$. The results of this fitting for field angle scans at the three different temperatures are shown in Fig. 3, with $q$ normalized to the value at $H_{\phi} = 90$ degrees ($q_0$) for each dataset. \

The behaviour of $q$ as a function of field angle is shown in Fig.~3a. A clear difference between the high temperature and low temperature regimes is observed. In the 50 K dataset $q$ initially decreases slightly as the applied field is rotated from the \hkl[110] to \hkl[1 1 -1] direction and then increases as the field is rotated further, up to a maximum when H is along \hkl[00-1]. This behaviour is reversed at low temperatures (both at 12~K and 5~K), where the magnitude of the conical wavevector instead increases to a maximum when the field is rotated from \hkl[110] to \hkl[1 1 -1], before quickly decreasing as the field angle approaches the \hkl[00-1] direction. In Fig. 3b, the effects of the anisotropic interactions can also clearly be seen to be more significant in the  low-temperature datasets, as in them $\phi$ deviates substantially from linearity after the magnetic field passes through \hkl[1 1 -1], as compared to the 50 K dataset where $\phi$ deviates only slightly from the expected linear trend in the absence of anisotropic interactions. \

The dependence of $q$ on $H_{\phi}$ allows us to decouple the two dominant anisotropic interactions. We start by using the non-constant terms within the free energy expansion derived by Bak and Jensen\cite{Bak1980}, which is valid for systems of $P$2$_{\textnormal{1}}$3 crystal symmetry that host slowly-varying magnetization densities $\mathbf{m}(\mathbf{r})$\cite{Landau1980}:

\begin{equation} \label{eq1}
\begin{split}
F(\mathbf{r}) &=  D\mathbf{m}(\mathbf{r})\cdot(\nabla\times\mathbf{m}(\mathbf{r})) \\
 &+\frac{1}{2}A[(\nabla m_x)^2+(\nabla m_y)^2+(\nabla m_z)^2] \\
 &+ \frac{1}{2}\gamma[(\frac{\partial m_x}{\partial x})^2+(\frac{\partial m_y}{\partial y})^2+(\frac{\partial m_z}{\partial z})^2]\\
 &+ K(m_x^4 + m_y^4+m_z^4)
 \end{split}
\end{equation}

Where $D, A$ are the familiar Dzyaloshinskii-Moriya and exchange stiffness, and $K, \gamma$ are the 4$^\textnormal{th}$ order MCA and AEI constants respectively.  In the case of $D = 0$, and $A, \gamma>0$, the spin-texture reduces to a simple ferromagnet, with the spin direction given by the sign of the MCA constant $K$. We approximate a model spin-texture using a conical ansatz:

\begin{equation} \label{eq2}
\mathbf{m}(\mathbf{r})/M_s = \sin\theta(\cos(\mathbf{q}\cdot\mathbf{r})\uvec{e}_1+\sin(\mathbf{q}\cdot\mathbf{r})\uvec{e}_2) + \cos\theta\uvec{e}_3
\end{equation}
where we follow the convention that $\theta$ is the conical angle, and $\{\uvec{e}_n\}$ define three mutually orthogonal basis vectors, with $\uvec{e}_3 \parallel \mathbf{q}$.  To compare with our experimental observations, we consider a conical texture with the wavevector restricted to the plane spanned by the three high cubic-symmetry directions, such that $\mathbf{q} = (\frac{q}{\sqrt2}\sin\phi, \frac{q}{\sqrt2}\sin\phi, q\cos\phi)$. Using this form for the magnetic wavevector and substituting (\ref{eq2}) into (\ref{eq1}), integrating over one conical period, $\lambda$, and differentiating with respect to $q$ we find a single stable solution:
\begin{equation} \label{eq4}
\begin{split}
\frac{\partial}{\partial q}\overline{F}(q, \phi) &= \sin^2\theta[D + Aq + \frac{1}{2}\gamma q \sin^2\phi(3\cos^2\phi + 1)] \\
q &= \frac{|D|}{A + \frac{1}{2}\gamma\sin^2\phi(3\cos^2\phi+1)}
\end{split}
\end{equation}
The implications of this result can be seen in Fig. 4(a-c), which show the experimentally measured values of the conical wavevector $q$ for different conical wavevector angles, $\phi$, at 50, 12 and 5 K respectively. At 50 K, the increased wavevector at \hkl[00-1] compared to \hkl[11-1] is a clear indication of a positive exchange anisotropy constant $\gamma$, where energy costs for directions other than the \hkl<100>, are compensated by a shrinking of the conical wavevector. This behaviour reverses with decreasing temperature, shown in Fig.~4b and c, where the wavevector maxima are located at the \hkl[11-1] and minima towards the \hkl[00-1], suggesting a change of sign of $\gamma$. \

\begin{figure}
\begin{center}
    \centering
    \begin{overpic}[width=0.99\linewidth]{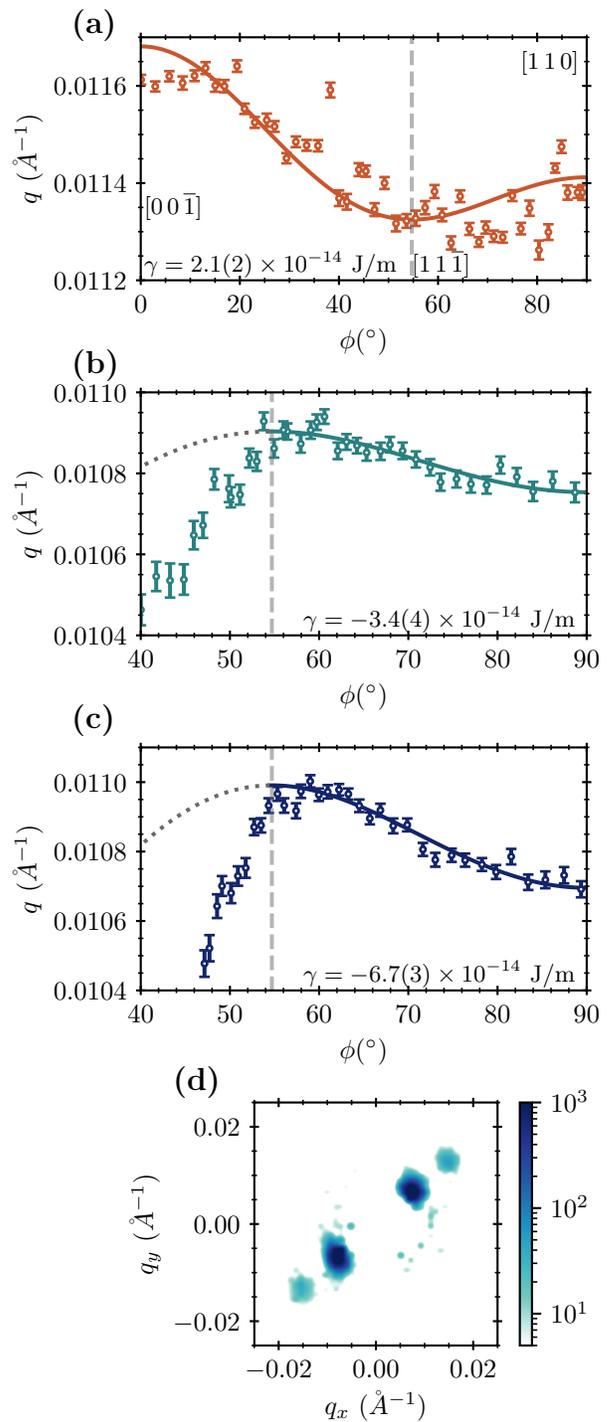}
    \put(6,98){\large\textbf{(a)}}
    \put(6,74.5){\large\textbf{(b)}}
    \put(6,50){\large\textbf{(c)}}
    \put(13,25){\large\textbf{(d)}}
    \put(37,95.5){\hkl[110]}
    \put(29.5,81.5){\hkl[11-1]}
    \put(11,85.5){\hkl[00-1]}
    \put(11.0,81.5){$\gamma = 2.1(2)\times 10^{-14}$ \si{J/m}}
    \put(22,57){$\gamma =- 3.4(4)\times 10^{-14}$ \si{J/m}}
    \put(22,32.5){$\gamma =- 6.7(3)\times 10^{-14}$ \si{J/m}}
	\end{overpic}    
    \caption{\small Values of the extracted wavevector magnitude ($q$) against wavevector angle ($\phi$) with (a-c) being the 5, 12 and 50 K datasets respectively. Fits to Eq. 4 are shown by the bold lines, with dashed lines showing data ignored in the fitting procedure due to the appearance of higher-order peaks, shown in (d). Values of $\gamma$ shown in panels using a value of  $A = 4.4\times 10^{-13}$ \si{J/m}.}
    \label{fig:2Dfits}
\end{center}
\end{figure}

Using the experimentally determined value for the exchange stiffness,  $A = 4.4\times 10^{-13}$ \si{J/m} from $T_\textnormal{C} = 57$ K \cite{Moody2018, White2007}, and fitting (3) to the data in Fig.~4, we find $\gamma = 2.1(2)\times 10^{-14}$ \si{J/m} at 50 K, $- 3.4(4)\times 10^{-14}$ at 12~K, and $- 6.7(3)\times 10^{-14}$ \si{J/m}  at 5 K. The low-temperature values are consistent with theoretical values of $\gamma$ required for TC formation \cite{Qian2018}, in agreement with the TC states observed in this study.

During the refinement of the low temperature datasets, only angles above 55 degrees were used. This is because at low temperatures, the direction of the conical wavevector deviates significantly from the magnetic field direction for $\phi$ $<$ 55 degrees, inducing a component of the magnetic field perpendicular to the conical wavevector. Applying transverse fields to a helical state with a pinned \textbf{q} direction is known to deform the helix \cite{Izyumov_1984}, introducing higher-order components to the Fourier transform of the magnetic state. The presence of higher order peaks in our SANS pattern shown in Fig. 4d suggest that similar deformations are occurring in our tilted conical state. For helices, these deformations are known to reduce the value of $q$ as the perpendicular field strength increases, which would be consistent with our observations of a large q discernibly where the higher order peaks are prominent. Fully accounting for these deformations within our model spin-texture remains an avenue intended for future study.

As shown in Fig.~1b(i), in the absence of cubic anisotropy, a change of sign of the AEI from positive to negative with decreasing temperature would induce a helical reorientation from the \hkl<100> to \hkl<111> directions. In order for the orientation of the helical wavevector to remain aligned along the \hkl<1 0 0> directions(consistent with experimental observations), we find that $K$ must be negative, with the condition $|K|>\frac{4}{3}|\gamma| q^2$. This requires that the magnitude of $K$ increases with decreasing temperatures. This increase in $|K|$ at low temperatures is also in agreement with previous work discussing the origin of the low temperature skyrmion state \cite{Halder2018}. \

In conclusion, we performed SANS on a single crystal of \ce{(Cu_{0.98}Zn_{0.02})_2OSeO3}, using a novel 3D vector magnet to decouple the anisotropic interactions within the material by investigating the behaviour of the magnetic conical state with a rotating magnetic field of constant magnitude. We observed a change of behaviour of the magnitude of the conical wavevector, $q$, as a function of wavevector angle, $\phi$, whereby the crystal directions corresponding to a maximum or minimum in $q$ reversed with cooling from 50 K to 12 K. We have explain this using a mean-field model and find that the AEI changes sign between these temperatures, with fitted values consistent with those required for tilted conical state formation. Unlike in the related compound \ce{FeGe}, helical reorientation does not occur within \ce{Cu2OSeO3} with the change in sign of the AEI. This is due to an increasingly negative cubic anisotropy, which increases in magnitude faster than the AEI, causing the helical wavevectors to remain along the \hkl<100> directions. We believe our finding that the AEI changes sign at similar temperatures to the occurrence of the LTS and TC magnetic phases will be highly useful for  understanding the formation and stabilisation of these newly discovered magnetic textures within this material.

\section{acknowledgements}
This work was supported by the UK Skyrmion Project EPSRC Programme Grant (No. EP/N032128/1). The SANS experiment at the
ISIS Pulsed Neutron and Muon Source was supported by a
beamtime allocation from the Science and Technology Facilities Council, Proposal No. 1920501. M.N.W. acknowledges
the support of the Natural Sciences and Engineering Research
Council of Canada (NSERC).
\bibliography{bib}

\begin{thebibliography}{43}%
\makeatletter
\providecommand \@ifxundefined [1]{%
 \@ifx{#1\undefined}
}%
\providecommand \@ifnum [1]{%
 \ifnum #1\expandafter \@firstoftwo
 \else \expandafter \@secondoftwo
 \fi
}%
\providecommand \@ifx [1]{%
 \ifx #1\expandafter \@firstoftwo
 \else \expandafter \@secondoftwo
 \fi
}%
\providecommand \natexlab [1]{#1}%
\providecommand \enquote  [1]{``#1''}%
\providecommand \bibnamefont  [1]{#1}%
\providecommand \bibfnamefont [1]{#1}%
\providecommand \citenamefont [1]{#1}%
\providecommand \href@noop [0]{\@secondoftwo}%
\providecommand \href [0]{\begingroup \@sanitize@url \@href}%
\providecommand \@href[1]{\@@startlink{#1}\@@href}%
\providecommand \@@href[1]{\endgroup#1\@@endlink}%
\providecommand \@sanitize@url [0]{\catcode `\\12\catcode `\$12\catcode
  `\&12\catcode `\#12\catcode `\^12\catcode `\_12\catcode `\%12\relax}%
\providecommand \@@startlink[1]{}%
\providecommand \@@endlink[0]{}%
\providecommand \url  [0]{\begingroup\@sanitize@url \@url }%
\providecommand \@url [1]{\endgroup\@href {#1}{\urlprefix }}%
\providecommand \urlprefix  [0]{URL }%
\providecommand \Eprint [0]{\href }%
\providecommand \doibase [0]{https://doi.org/}%
\providecommand \selectlanguage [0]{\@gobble}%
\providecommand \bibinfo  [0]{\@secondoftwo}%
\providecommand \bibfield  [0]{\@secondoftwo}%
\providecommand \translation [1]{[#1]}%
\providecommand \BibitemOpen [0]{}%
\providecommand \bibitemStop [0]{}%
\providecommand \bibitemNoStop [0]{.\EOS\space}%
\providecommand \EOS [0]{\spacefactor3000\relax}%
\providecommand \BibitemShut  [1]{\csname bibitem#1\endcsname}%
\let\auto@bib@innerbib\@empty
\bibitem [{\citenamefont {Lancaster}(2019)}]{Lancaster2019}%
  \BibitemOpen
  \bibfield  {author} {\bibinfo {author} {\bibfnamefont {T.}~\bibnamefont
  {Lancaster}},\ }\bibfield  {title} {\bibinfo {title} {Skyrmions in magnetic
  materials},\ }\href {https://doi.org/10.1080/00107514.2019.1699352}
  {\bibfield  {journal} {\bibinfo  {journal} {Contemporary Physics}\ }\textbf
  {\bibinfo {volume} {60}},\ \bibinfo {pages} {246} (\bibinfo {year}
  {2019})}\BibitemShut {NoStop}%
\bibitem [{\citenamefont {Dzyaloshinskii}(1964)}]{Dzya1964}%
  \BibitemOpen
  \bibfield  {author} {\bibinfo {author} {\bibfnamefont {I.~E.}\ \bibnamefont
  {Dzyaloshinskii}},\ }\bibfield  {title} {\bibinfo {title} {Theory of
  helicoidal structures in antiferromagnets},\ }\href@noop {} {\bibfield
  {journal} {\bibinfo  {journal} {J. Exptl. Theoret. Phys.}\ }\textbf {\bibinfo
  {volume} {46}},\ \bibinfo {pages} {1420} (\bibinfo {year}
  {1964})}\BibitemShut {NoStop}%
\bibitem [{\citenamefont {Neubauer}\ \emph {et~al.}(2009)\citenamefont
  {Neubauer}, \citenamefont {Pfleiderer}, \citenamefont {Binz}, \citenamefont
  {Rosch}, \citenamefont {Ritz}, \citenamefont {Niklowitz},\ and\ \citenamefont
  {B\"oni}}]{Neubauer2009}%
  \BibitemOpen
  \bibfield  {author} {\bibinfo {author} {\bibfnamefont {A.}~\bibnamefont
  {Neubauer}}, \bibinfo {author} {\bibfnamefont {C.}~\bibnamefont
  {Pfleiderer}}, \bibinfo {author} {\bibfnamefont {B.}~\bibnamefont {Binz}},
  \bibinfo {author} {\bibfnamefont {A.}~\bibnamefont {Rosch}}, \bibinfo
  {author} {\bibfnamefont {R.}~\bibnamefont {Ritz}}, \bibinfo {author}
  {\bibfnamefont {P.~G.}\ \bibnamefont {Niklowitz}},\ and\ \bibinfo {author}
  {\bibfnamefont {P.}~\bibnamefont {B\"oni}},\ }\bibfield  {title} {\bibinfo
  {title} {Topological \ce{H}all effect in the {A} phase of \ce{MnSi}},\ }\href
  {https://doi.org/10.1103/PhysRevLett.102.186602} {\bibfield  {journal}
  {\bibinfo  {journal} {Phys. Rev. Lett.}\ }\textbf {\bibinfo {volume} {102}},\
  \bibinfo {pages} {186602} (\bibinfo {year} {2009})}\BibitemShut {NoStop}%
\bibitem [{\citenamefont {Tanigaki}\ \emph {et~al.}(2015)\citenamefont
  {Tanigaki}, \citenamefont {Shibata}, \citenamefont {Kanazawa}, \citenamefont
  {Yu}, \citenamefont {Onose}, \citenamefont {Park}, \citenamefont {Shindo},\
  and\ \citenamefont {Tokura}}]{Tanigaki2015}%
  \BibitemOpen
  \bibfield  {author} {\bibinfo {author} {\bibfnamefont {T.}~\bibnamefont
  {Tanigaki}}, \bibinfo {author} {\bibfnamefont {K.}~\bibnamefont {Shibata}},
  \bibinfo {author} {\bibfnamefont {N.}~\bibnamefont {Kanazawa}}, \bibinfo
  {author} {\bibfnamefont {X.}~\bibnamefont {Yu}}, \bibinfo {author}
  {\bibfnamefont {Y.}~\bibnamefont {Onose}}, \bibinfo {author} {\bibfnamefont
  {H.~S.}\ \bibnamefont {Park}}, \bibinfo {author} {\bibfnamefont
  {D.}~\bibnamefont {Shindo}},\ and\ \bibinfo {author} {\bibfnamefont
  {Y.}~\bibnamefont {Tokura}},\ }\bibfield  {title} {\bibinfo {title}
  {Real-space observation of short-period cubic lattice of skyrmions in mnge},\
  }\href {https://doi.org/10.1021/acs.nanolett.5b02653} {\bibfield  {journal}
  {\bibinfo  {journal} {Nano Letters}\ }\textbf {\bibinfo {volume} {15}},\
  \bibinfo {pages} {5438} (\bibinfo {year} {2015})}\BibitemShut {NoStop}%
\bibitem [{\citenamefont {Wilhelm}\ \emph {et~al.}(2011)\citenamefont
  {Wilhelm}, \citenamefont {Baenitz}, \citenamefont {Schmidt}, \citenamefont
  {R\"o\ss{}ler}, \citenamefont {Leonov},\ and\ \citenamefont
  {Bogdanov}}]{Wilhelm2011}%
  \BibitemOpen
  \bibfield  {author} {\bibinfo {author} {\bibfnamefont {H.}~\bibnamefont
  {Wilhelm}}, \bibinfo {author} {\bibfnamefont {M.}~\bibnamefont {Baenitz}},
  \bibinfo {author} {\bibfnamefont {M.}~\bibnamefont {Schmidt}}, \bibinfo
  {author} {\bibfnamefont {U.~K.}\ \bibnamefont {R\"o\ss{}ler}}, \bibinfo
  {author} {\bibfnamefont {A.~A.}\ \bibnamefont {Leonov}},\ and\ \bibinfo
  {author} {\bibfnamefont {A.~N.}\ \bibnamefont {Bogdanov}},\ }\bibfield
  {title} {\bibinfo {title} {Precursor phenomena at the magnetic ordering of
  the cubic helimagnet \ce{FeGe}},\ }\href
  {https://doi.org/10.1103/PhysRevLett.107.127203} {\bibfield  {journal}
  {\bibinfo  {journal} {Phys. Rev. Lett.}\ }\textbf {\bibinfo {volume} {107}},\
  \bibinfo {pages} {127203} (\bibinfo {year} {2011})}\BibitemShut {NoStop}%
\bibitem [{\citenamefont {Yu}\ \emph {et~al.}(2011)\citenamefont {Yu},
  \citenamefont {Kanazawa}, \citenamefont {Onose}, \citenamefont {Kimoto},
  \citenamefont {Zhang}, \citenamefont {Ishiwata}, \citenamefont {Matsui},\
  and\ \citenamefont {Tokura}}]{Yu2011}%
  \BibitemOpen
  \bibfield  {author} {\bibinfo {author} {\bibfnamefont {X.~Z.}\ \bibnamefont
  {Yu}}, \bibinfo {author} {\bibfnamefont {N.}~\bibnamefont {Kanazawa}},
  \bibinfo {author} {\bibfnamefont {Y.}~\bibnamefont {Onose}}, \bibinfo
  {author} {\bibfnamefont {K.}~\bibnamefont {Kimoto}}, \bibinfo {author}
  {\bibfnamefont {W.~Z.}\ \bibnamefont {Zhang}}, \bibinfo {author}
  {\bibfnamefont {S.}~\bibnamefont {Ishiwata}}, \bibinfo {author}
  {\bibfnamefont {Y.}~\bibnamefont {Matsui}},\ and\ \bibinfo {author}
  {\bibfnamefont {Y.}~\bibnamefont {Tokura}},\ }\bibfield  {title} {\bibinfo
  {title} {Near room-temperature formation of a skyrmion crystal in thin-films
  of the helimagnet \ce{FeGe}},\ }\href {https://doi.org/10.1038/nmat2916}
  {\bibfield  {journal} {\bibinfo  {journal} {Nature Materials}\ }\textbf
  {\bibinfo {volume} {10}},\ \bibinfo {pages} {106} (\bibinfo {year}
  {2011})}\BibitemShut {NoStop}%
\bibitem [{\citenamefont {Seki}\ \emph {et~al.}(2012)\citenamefont {Seki},
  \citenamefont {Yu}, \citenamefont {Ishiwata},\ and\ \citenamefont
  {Tokura}}]{Seki2012}%
  \BibitemOpen
  \bibfield  {author} {\bibinfo {author} {\bibfnamefont {S.}~\bibnamefont
  {Seki}}, \bibinfo {author} {\bibfnamefont {X.~Z.}\ \bibnamefont {Yu}},
  \bibinfo {author} {\bibfnamefont {S.}~\bibnamefont {Ishiwata}},\ and\
  \bibinfo {author} {\bibfnamefont {Y.}~\bibnamefont {Tokura}},\ }\bibfield
  {title} {\bibinfo {title} {Observation of skyrmions in a multiferroic
  material},\ }\href {https://doi.org/10.1126/science.1214143} {\bibfield
  {journal} {\bibinfo  {journal} {Science}\ }\textbf {\bibinfo {volume}
  {336}},\ \bibinfo {pages} {198} (\bibinfo {year} {2012})}\BibitemShut
  {NoStop}%
\bibitem [{\citenamefont {Tokunaga}\ \emph {et~al.}(2015)\citenamefont
  {Tokunaga}, \citenamefont {Yu}, \citenamefont {White}, \citenamefont
  {R{\o}nnow}, \citenamefont {Morikawa}, \citenamefont {Taguchi},\ and\
  \citenamefont {Tokura}}]{Tokunaga2015}%
  \BibitemOpen
  \bibfield  {author} {\bibinfo {author} {\bibfnamefont {Y.}~\bibnamefont
  {Tokunaga}}, \bibinfo {author} {\bibfnamefont {X.~Z.}\ \bibnamefont {Yu}},
  \bibinfo {author} {\bibfnamefont {J.~S.}\ \bibnamefont {White}}, \bibinfo
  {author} {\bibfnamefont {H.~M.}\ \bibnamefont {R{\o}nnow}}, \bibinfo {author}
  {\bibfnamefont {D.}~\bibnamefont {Morikawa}}, \bibinfo {author}
  {\bibfnamefont {Y.}~\bibnamefont {Taguchi}},\ and\ \bibinfo {author}
  {\bibfnamefont {Y.}~\bibnamefont {Tokura}},\ }\bibfield  {title} {\bibinfo
  {title} {A new class of chiral materials hosting magnetic skyrmions beyond
  room temperature},\ }\href {https://doi.org/10.1038/ncomms8638} {\bibfield
  {journal} {\bibinfo  {journal} {Nature Communications}\ }\textbf {\bibinfo
  {volume} {6}},\ \bibinfo {pages} {7638} (\bibinfo {year} {2015})}\BibitemShut
  {NoStop}%
\bibitem [{\citenamefont {Blundell}(2001)}]{Blundell}%
  \BibitemOpen
  \bibfield  {author} {\bibinfo {author} {\bibfnamefont {S.}~\bibnamefont
  {Blundell}},\ }\href@noop {} {\emph {\bibinfo {title} {Magnetism in Condensed
  Matter}}}\ (\bibinfo  {publisher} {Oxford University Press},\ \bibinfo
  {address} {Oxford},\ \bibinfo {year} {2001})\BibitemShut {NoStop}%
\bibitem [{\citenamefont {Ishikawa}\ \emph {et~al.}(1976)\citenamefont
  {Ishikawa}, \citenamefont {Tajima}, \citenamefont {Bloch},\ and\
  \citenamefont {Roth}}]{ISHIKAWA1976}%
  \BibitemOpen
  \bibfield  {author} {\bibinfo {author} {\bibfnamefont {Y.}~\bibnamefont
  {Ishikawa}}, \bibinfo {author} {\bibfnamefont {K.}~\bibnamefont {Tajima}},
  \bibinfo {author} {\bibfnamefont {D.}~\bibnamefont {Bloch}},\ and\ \bibinfo
  {author} {\bibfnamefont {M.}~\bibnamefont {Roth}},\ }\bibfield  {title}
  {\bibinfo {title} {Helical spin structure in manganese silicide \ce{MnSi}},\
  }\href@noop {} {\bibfield  {journal} {\bibinfo  {journal} {Solid State
  Communications}\ }\textbf {\bibinfo {volume} {19}},\ \bibinfo {pages} {525}
  (\bibinfo {year} {1976})}\BibitemShut {NoStop}%
\bibitem [{\citenamefont {{Bak}}\ and\ \citenamefont
  {{Jensen}}(1980)}]{Bak1980}%
  \BibitemOpen
  \bibfield  {author} {\bibinfo {author} {\bibfnamefont {P.}~\bibnamefont
  {{Bak}}}\ and\ \bibinfo {author} {\bibfnamefont {M.~H.}\ \bibnamefont
  {{Jensen}}},\ }\bibfield  {title} {\bibinfo {title} {{Theory of helical
  magnetic structures and phase transitions in \ce{MnSi} and \ce{FeGe}}},\
  }\href {https://doi.org/10.1088/0022-3719/13/31/002} {\bibfield  {journal}
  {\bibinfo  {journal} {Journal of Physics C Solid State Physics}\ }\textbf
  {\bibinfo {volume} {13}},\ \bibinfo {pages} {L881} (\bibinfo {year}
  {1980})}\BibitemShut {NoStop}%
\bibitem [{\citenamefont {Nakanishi}\ \emph {et~al.}(1980)\citenamefont
  {Nakanishi}, \citenamefont {Yanase}, \citenamefont {Hasegawa},\ and\
  \citenamefont {Kataoka}}]{Nakanishi1980}%
  \BibitemOpen
  \bibfield  {author} {\bibinfo {author} {\bibfnamefont {O.}~\bibnamefont
  {Nakanishi}}, \bibinfo {author} {\bibfnamefont {A.}~\bibnamefont {Yanase}},
  \bibinfo {author} {\bibfnamefont {A.}~\bibnamefont {Hasegawa}},\ and\
  \bibinfo {author} {\bibfnamefont {M.}~\bibnamefont {Kataoka}},\ }\bibfield
  {title} {\bibinfo {title} {The origin of the helical spin density wave in
  \ce{MnSi}},\ }\href@noop {} {\bibfield  {journal} {\bibinfo  {journal} {Solid
  State Communications}\ }\textbf {\bibinfo {volume} {35}},\ \bibinfo {pages}
  {995} (\bibinfo {year} {1980})}\BibitemShut {NoStop}%
\bibitem [{\citenamefont {Maleyev}(2006)}]{Maleyev2006}%
  \BibitemOpen
  \bibfield  {author} {\bibinfo {author} {\bibfnamefont {S.~V.}\ \bibnamefont
  {Maleyev}},\ }\bibfield  {title} {\bibinfo {title} {Cubic magnets with
  {Dzyaloshinskii}-{Moriya} interaction at low temperature},\ }\href
  {https://doi.org/10.1103/PhysRevB.73.174402} {\bibfield  {journal} {\bibinfo
  {journal} {Phys. Rev. B}\ }\textbf {\bibinfo {volume} {73}},\ \bibinfo
  {pages} {174402} (\bibinfo {year} {2006})}\BibitemShut {NoStop}%
\bibitem [{\citenamefont {Ukleev}\ \emph {et~al.}(2021)\citenamefont {Ukleev},
  \citenamefont {Utesov}, \citenamefont {Yu}, \citenamefont {Luo},
  \citenamefont {Chen}, \citenamefont {Radu}, \citenamefont {Yamasaki},
  \citenamefont {Kanazawa}, \citenamefont {Tokura}, \citenamefont {Arima},\
  and\ \citenamefont {White}}]{Ukleev2021}%
  \BibitemOpen
  \bibfield  {author} {\bibinfo {author} {\bibfnamefont {V.}~\bibnamefont
  {Ukleev}}, \bibinfo {author} {\bibfnamefont {O.}~\bibnamefont {Utesov}},
  \bibinfo {author} {\bibfnamefont {L.}~\bibnamefont {Yu}}, \bibinfo {author}
  {\bibfnamefont {C.}~\bibnamefont {Luo}}, \bibinfo {author} {\bibfnamefont
  {K.}~\bibnamefont {Chen}}, \bibinfo {author} {\bibfnamefont {F.}~\bibnamefont
  {Radu}}, \bibinfo {author} {\bibfnamefont {Y.}~\bibnamefont {Yamasaki}},
  \bibinfo {author} {\bibfnamefont {N.}~\bibnamefont {Kanazawa}}, \bibinfo
  {author} {\bibfnamefont {Y.}~\bibnamefont {Tokura}}, \bibinfo {author}
  {\bibfnamefont {T.-h.}\ \bibnamefont {Arima}},\ and\ \bibinfo {author}
  {\bibfnamefont {J.~S.}\ \bibnamefont {White}},\ }\bibfield  {title} {\bibinfo
  {title} {Signature of anisotropic exchange interaction revealed by
  vector-field control of the helical order in a \ce{FeGe} thin plate},\ }\href
  {https://doi.org/10.1103/PhysRevResearch.3.013094} {\bibfield  {journal}
  {\bibinfo  {journal} {Phys. Rev. Research}\ }\textbf {\bibinfo {volume}
  {3}},\ \bibinfo {pages} {013094} (\bibinfo {year} {2021})}\BibitemShut
  {NoStop}%
\bibitem [{\citenamefont {Lebech}\ \emph {et~al.}(1989)\citenamefont {Lebech},
  \citenamefont {Bernhard},\ and\ \citenamefont {Freltoft}}]{Lebech_1989}%
  \BibitemOpen
  \bibfield  {author} {\bibinfo {author} {\bibfnamefont {B.}~\bibnamefont
  {Lebech}}, \bibinfo {author} {\bibfnamefont {J.}~\bibnamefont {Bernhard}},\
  and\ \bibinfo {author} {\bibfnamefont {T.}~\bibnamefont {Freltoft}},\
  }\bibfield  {title} {\bibinfo {title} {Magnetic structures of cubic \ce{FeGe}
  studied by small-angle neutron scattering},\ }\href@noop {} {\bibfield
  {journal} {\bibinfo  {journal} {Journal of Physics: Condensed Matter}\
  }\textbf {\bibinfo {volume} {1}},\ \bibinfo {pages} {6105} (\bibinfo {year}
  {1989})}\BibitemShut {NoStop}%
\bibitem [{\citenamefont {Plumer}(1990)}]{Plumer_1990}%
  \BibitemOpen
  \bibfield  {author} {\bibinfo {author} {\bibfnamefont {M.~L.}\ \bibnamefont
  {Plumer}},\ }\bibfield  {title} {\bibinfo {title} {Wavevector and spin-flop
  transitions in cubic \ce{FeGe}},\ }\href@noop {} {\bibfield  {journal}
  {\bibinfo  {journal} {Journal of Physics: Condensed Matter}\ }\textbf
  {\bibinfo {volume} {2}},\ \bibinfo {pages} {7503} (\bibinfo {year}
  {1990})}\BibitemShut {NoStop}%
\bibitem [{\citenamefont {Ishikawa}\ \emph {et~al.}(1977)\citenamefont
  {Ishikawa}, \citenamefont {Komatsubara},\ and\ \citenamefont
  {Bloch}}]{ISHIKAWA1977}%
  \BibitemOpen
  \bibfield  {author} {\bibinfo {author} {\bibfnamefont {Y.}~\bibnamefont
  {Ishikawa}}, \bibinfo {author} {\bibfnamefont {T.}~\bibnamefont
  {Komatsubara}},\ and\ \bibinfo {author} {\bibfnamefont {D.}~\bibnamefont
  {Bloch}},\ }\bibfield  {title} {\bibinfo {title} {Magnetic phase diagram of
  \ce{MnSi}},\ }\href@noop {} {\bibfield  {journal} {\bibinfo  {journal}
  {Physica B}\ }\textbf {\bibinfo {volume} {86-88}},\ \bibinfo {pages} {401}
  (\bibinfo {year} {1977})}\BibitemShut {NoStop}%
\bibitem [{\citenamefont {Ishimoto}\ \emph {et~al.}(1990)\citenamefont
  {Ishimoto}, \citenamefont {Yamauchi}, \citenamefont {Yamaguchi},
  \citenamefont {Suzuki}, \citenamefont {Arai}, \citenamefont {Furusaka},\ and\
  \citenamefont {Endoh}}]{ISHIMOTO1990}%
  \BibitemOpen
  \bibfield  {author} {\bibinfo {author} {\bibfnamefont {K.}~\bibnamefont
  {Ishimoto}}, \bibinfo {author} {\bibfnamefont {H.}~\bibnamefont {Yamauchi}},
  \bibinfo {author} {\bibfnamefont {Y.}~\bibnamefont {Yamaguchi}}, \bibinfo
  {author} {\bibfnamefont {J.}~\bibnamefont {Suzuki}}, \bibinfo {author}
  {\bibfnamefont {M.}~\bibnamefont {Arai}}, \bibinfo {author} {\bibfnamefont
  {M.}~\bibnamefont {Furusaka}},\ and\ \bibinfo {author} {\bibfnamefont
  {Y.}~\bibnamefont {Endoh}},\ }\bibfield  {title} {\bibinfo {title} {Anomalous
  region in the magnetic phase diagram of (\ce{Fe},\ce{Co})\ce{Si}},\
  }\href@noop {} {\bibfield  {journal} {\bibinfo  {journal} {Journal of
  Magnetism and Magnetic Materials}\ }\textbf {\bibinfo {volume} {90-91}},\
  \bibinfo {pages} {163} (\bibinfo {year} {1990})}\BibitemShut {NoStop}%
\bibitem [{\citenamefont {M{\"u}hlbauer}\ \emph {et~al.}(2009)\citenamefont
  {M{\"u}hlbauer}, \citenamefont {Binz}, \citenamefont {Jonietz}, \citenamefont
  {Pfleiderer}, \citenamefont {Rosch}, \citenamefont {Neubauer}, \citenamefont
  {Georgii},\ and\ \citenamefont {B{\"o}ni}}]{Muhlbauer2009}%
  \BibitemOpen
  \bibfield  {author} {\bibinfo {author} {\bibfnamefont {S.}~\bibnamefont
  {M{\"u}hlbauer}}, \bibinfo {author} {\bibfnamefont {B.}~\bibnamefont {Binz}},
  \bibinfo {author} {\bibfnamefont {F.}~\bibnamefont {Jonietz}}, \bibinfo
  {author} {\bibfnamefont {C.}~\bibnamefont {Pfleiderer}}, \bibinfo {author}
  {\bibfnamefont {A.}~\bibnamefont {Rosch}}, \bibinfo {author} {\bibfnamefont
  {A.}~\bibnamefont {Neubauer}}, \bibinfo {author} {\bibfnamefont
  {R.}~\bibnamefont {Georgii}},\ and\ \bibinfo {author} {\bibfnamefont
  {P.}~\bibnamefont {B{\"o}ni}},\ }\bibfield  {title} {\bibinfo {title}
  {Skyrmion lattice in a chiral magnet},\ }\href
  {https://doi.org/10.1126/science.1166767} {\bibfield  {journal} {\bibinfo
  {journal} {Science}\ }\textbf {\bibinfo {volume} {323}},\ \bibinfo {pages}
  {915} (\bibinfo {year} {2009})}\BibitemShut {NoStop}%
\bibitem [{\citenamefont {R{\"o}{\ss}ler}\ \emph {et~al.}(2006)\citenamefont
  {R{\"o}{\ss}ler}, \citenamefont {Bogdanov},\ and\ \citenamefont
  {Pfleiderer}}]{Robler2006}%
  \BibitemOpen
  \bibfield  {author} {\bibinfo {author} {\bibfnamefont {U.~K.}\ \bibnamefont
  {R{\"o}{\ss}ler}}, \bibinfo {author} {\bibfnamefont {A.~N.}\ \bibnamefont
  {Bogdanov}},\ and\ \bibinfo {author} {\bibfnamefont {C.}~\bibnamefont
  {Pfleiderer}},\ }\bibfield  {title} {\bibinfo {title} {Spontaneous skyrmion
  ground states in magnetic metals},\ }\href@noop {} {\bibfield  {journal}
  {\bibinfo  {journal} {Nature}\ }\textbf {\bibinfo {volume} {442}},\ \bibinfo
  {pages} {797} (\bibinfo {year} {2006})}\BibitemShut {NoStop}%
\bibitem [{\citenamefont {Vousden}\ \emph {et~al.}(2016)\citenamefont
  {Vousden}, \citenamefont {Albert}, \citenamefont {Beg}, \citenamefont
  {Bisotti}, \citenamefont {Carey}, \citenamefont {Chernyshenko}, \citenamefont
  {Cortés-Ortuño}, \citenamefont {Wang}, \citenamefont {Hovorka},
  \citenamefont {Marrows},\ and\ \citenamefont {Fangohr}}]{Vousden2016}%
  \BibitemOpen
  \bibfield  {author} {\bibinfo {author} {\bibfnamefont {M.}~\bibnamefont
  {Vousden}}, \bibinfo {author} {\bibfnamefont {M.}~\bibnamefont {Albert}},
  \bibinfo {author} {\bibfnamefont {M.}~\bibnamefont {Beg}}, \bibinfo {author}
  {\bibfnamefont {M.-A.}\ \bibnamefont {Bisotti}}, \bibinfo {author}
  {\bibfnamefont {R.}~\bibnamefont {Carey}}, \bibinfo {author} {\bibfnamefont
  {D.}~\bibnamefont {Chernyshenko}}, \bibinfo {author} {\bibfnamefont
  {D.}~\bibnamefont {Cortés-Ortuño}}, \bibinfo {author} {\bibfnamefont
  {W.}~\bibnamefont {Wang}}, \bibinfo {author} {\bibfnamefont {O.}~\bibnamefont
  {Hovorka}}, \bibinfo {author} {\bibfnamefont {C.~H.}\ \bibnamefont
  {Marrows}},\ and\ \bibinfo {author} {\bibfnamefont {H.}~\bibnamefont
  {Fangohr}},\ }\bibfield  {title} {\bibinfo {title} {Skyrmions in thin films
  with easy-plane magnetocrystalline anisotropy},\ }\href
  {https://doi.org/10.1063/1.4945262} {\bibfield  {journal} {\bibinfo
  {journal} {Applied Physics Letters}\ }\textbf {\bibinfo {volume} {108}},\
  \bibinfo {pages} {132406} (\bibinfo {year} {2016})}\BibitemShut {NoStop}%
\bibitem [{\citenamefont {Bogdanov}\ and\ \citenamefont
  {Panagopoulos}(2020)}]{Bogdanov2020}%
  \BibitemOpen
  \bibfield  {author} {\bibinfo {author} {\bibfnamefont {A.~N.}\ \bibnamefont
  {Bogdanov}}\ and\ \bibinfo {author} {\bibfnamefont {C.}~\bibnamefont
  {Panagopoulos}},\ }\bibfield  {title} {\bibinfo {title} {Physical foundations
  and basic properties of magnetic skyrmions},\ }\href
  {https://doi.org/10.1038/s42254-020-0203-7} {\bibfield  {journal} {\bibinfo
  {journal} {Nature Reviews Physics}\ }\textbf {\bibinfo {volume} {2}},\
  \bibinfo {pages} {492} (\bibinfo {year} {2020})}\BibitemShut {NoStop}%
\bibitem [{\citenamefont {Zhang}\ \emph {et~al.}(2015)\citenamefont {Zhang},
  \citenamefont {Ezawa},\ and\ \citenamefont {Zhou}}]{Zhang2015}%
  \BibitemOpen
  \bibfield  {author} {\bibinfo {author} {\bibfnamefont {X.}~\bibnamefont
  {Zhang}}, \bibinfo {author} {\bibfnamefont {M.}~\bibnamefont {Ezawa}},\ and\
  \bibinfo {author} {\bibfnamefont {Y.}~\bibnamefont {Zhou}},\ }\bibfield
  {title} {\bibinfo {title} {Magnetic skyrmion logic gates: conversion,
  duplication and merging of skyrmions},\ }\href
  {https://doi.org/10.1038/srep09400} {\bibfield  {journal} {\bibinfo
  {journal} {Scientific Reports}\ }\textbf {\bibinfo {volume} {5}},\ \bibinfo
  {pages} {9400} (\bibinfo {year} {2015})}\BibitemShut {NoStop}%
\bibitem [{\citenamefont {Fert}\ \emph {et~al.}(2017)\citenamefont {Fert},
  \citenamefont {Reyren},\ and\ \citenamefont {Cros}}]{Fert2017}%
  \BibitemOpen
  \bibfield  {author} {\bibinfo {author} {\bibfnamefont {A.}~\bibnamefont
  {Fert}}, \bibinfo {author} {\bibfnamefont {N.}~\bibnamefont {Reyren}},\ and\
  \bibinfo {author} {\bibfnamefont {V.}~\bibnamefont {Cros}},\ }\bibfield
  {title} {\bibinfo {title} {Magnetic skyrmions: advances in physics and
  potential applications},\ }\href {https://doi.org/10.1038/natrevmats.2017.31}
  {\bibfield  {journal} {\bibinfo  {journal} {Nature Reviews Materials}\
  }\textbf {\bibinfo {volume} {2}},\ \bibinfo {pages} {17031} (\bibinfo {year}
  {2017})}\BibitemShut {NoStop}%
\bibitem [{\citenamefont {Zhang}\ \emph {et~al.}(2018)\citenamefont {Zhang},
  \citenamefont {Wang}, \citenamefont {Burn}, \citenamefont {Peng},
  \citenamefont {Berger}, \citenamefont {Bauer}, \citenamefont {Pfleiderer},
  \citenamefont {van~der Laan},\ and\ \citenamefont {Hesjedal}}]{Zhang2018}%
  \BibitemOpen
  \bibfield  {author} {\bibinfo {author} {\bibfnamefont {S.~L.}\ \bibnamefont
  {Zhang}}, \bibinfo {author} {\bibfnamefont {W.~W.}\ \bibnamefont {Wang}},
  \bibinfo {author} {\bibfnamefont {D.~M.}\ \bibnamefont {Burn}}, \bibinfo
  {author} {\bibfnamefont {H.}~\bibnamefont {Peng}}, \bibinfo {author}
  {\bibfnamefont {H.}~\bibnamefont {Berger}}, \bibinfo {author} {\bibfnamefont
  {A.}~\bibnamefont {Bauer}}, \bibinfo {author} {\bibfnamefont
  {C.}~\bibnamefont {Pfleiderer}}, \bibinfo {author} {\bibfnamefont
  {G.}~\bibnamefont {van~der Laan}},\ and\ \bibinfo {author} {\bibfnamefont
  {T.}~\bibnamefont {Hesjedal}},\ }\bibfield  {title} {\bibinfo {title}
  {Manipulation of skyrmion motion by magnetic field gradients},\ }\href
  {https://doi.org/10.1038/s41467-018-04563-4} {\bibfield  {journal} {\bibinfo
  {journal} {Nature Communications}\ }\textbf {\bibinfo {volume} {9}},\
  \bibinfo {pages} {2115} (\bibinfo {year} {2018})}\BibitemShut {NoStop}%
\bibitem [{\citenamefont {Sch{\"a}ffer}\ \emph {et~al.}(2019)\citenamefont
  {Sch{\"a}ffer}, \citenamefont {R{\'o}zsa}, \citenamefont {Berakdar},
  \citenamefont {Vedmedenko},\ and\ \citenamefont
  {Wiesendanger}}]{Schaffer2019}%
  \BibitemOpen
  \bibfield  {author} {\bibinfo {author} {\bibfnamefont {A.~F.}\ \bibnamefont
  {Sch{\"a}ffer}}, \bibinfo {author} {\bibfnamefont {L.}~\bibnamefont
  {R{\'o}zsa}}, \bibinfo {author} {\bibfnamefont {J.}~\bibnamefont {Berakdar}},
  \bibinfo {author} {\bibfnamefont {E.~Y.}\ \bibnamefont {Vedmedenko}},\ and\
  \bibinfo {author} {\bibfnamefont {R.}~\bibnamefont {Wiesendanger}},\
  }\bibfield  {title} {\bibinfo {title} {Stochastic dynamics and pattern
  formation of geometrically confined skyrmions},\ }\href
  {https://doi.org/10.1038/s42005-019-0176-y} {\bibfield  {journal} {\bibinfo
  {journal} {Communications Physics}\ }\textbf {\bibinfo {volume} {2}},\
  \bibinfo {pages} {72} (\bibinfo {year} {2019})}\BibitemShut {NoStop}%
\bibitem [{\citenamefont {Zhang}\ \emph {et~al.}(2020)\citenamefont {Zhang},
  \citenamefont {Zhu}, \citenamefont {Kang}, \citenamefont {Zhang},\ and\
  \citenamefont {Zhao}}]{Zhang2020}%
  \BibitemOpen
  \bibfield  {author} {\bibinfo {author} {\bibfnamefont {H.}~\bibnamefont
  {Zhang}}, \bibinfo {author} {\bibfnamefont {D.}~\bibnamefont {Zhu}}, \bibinfo
  {author} {\bibfnamefont {W.}~\bibnamefont {Kang}}, \bibinfo {author}
  {\bibfnamefont {Y.}~\bibnamefont {Zhang}},\ and\ \bibinfo {author}
  {\bibfnamefont {W.}~\bibnamefont {Zhao}},\ }\bibfield  {title} {\bibinfo
  {title} {Stochastic computing implemented by skyrmionic logic devices},\
  }\href {https://doi.org/10.1103/PhysRevApplied.13.054049} {\bibfield
  {journal} {\bibinfo  {journal} {Phys. Rev. Applied}\ }\textbf {\bibinfo
  {volume} {13}},\ \bibinfo {pages} {054049} (\bibinfo {year}
  {2020})}\BibitemShut {NoStop}%
\bibitem [{\citenamefont {Li}\ \emph {et~al.}(2017)\citenamefont {Li},
  \citenamefont {Kang}, \citenamefont {Huang}, \citenamefont {Zhang},
  \citenamefont {Zhou},\ and\ \citenamefont {Zhao}}]{Li2017}%
  \BibitemOpen
  \bibfield  {author} {\bibinfo {author} {\bibfnamefont {S.}~\bibnamefont
  {Li}}, \bibinfo {author} {\bibfnamefont {W.}~\bibnamefont {Kang}}, \bibinfo
  {author} {\bibfnamefont {Y.}~\bibnamefont {Huang}}, \bibinfo {author}
  {\bibfnamefont {X.}~\bibnamefont {Zhang}}, \bibinfo {author} {\bibfnamefont
  {Y.}~\bibnamefont {Zhou}},\ and\ \bibinfo {author} {\bibfnamefont
  {W.}~\bibnamefont {Zhao}},\ }\bibfield  {title} {\bibinfo {title} {Magnetic
  skyrmion-based artificial neuron device},\ }\href
  {https://doi.org/10.1088/1361-6528/aa7af5} {\bibfield  {journal} {\bibinfo
  {journal} {Nanotechnology}\ }\textbf {\bibinfo {volume} {28}},\ \bibinfo
  {pages} {31LT01} (\bibinfo {year} {2017})}\BibitemShut {NoStop}%
\bibitem [{\citenamefont {Song}\ \emph {et~al.}(2020)\citenamefont {Song},
  \citenamefont {Jeong}, \citenamefont {Pan}, \citenamefont {Zhang},
  \citenamefont {Xia}, \citenamefont {Cha}, \citenamefont {Park}, \citenamefont
  {Kim}, \citenamefont {Finizio}, \citenamefont {Raabe}, \citenamefont {Chang},
  \citenamefont {Zhou}, \citenamefont {Zhao}, \citenamefont {Kang},
  \citenamefont {Ju},\ and\ \citenamefont {Woo}}]{Song2020}%
  \BibitemOpen
  \bibfield  {author} {\bibinfo {author} {\bibfnamefont {K.~M.}\ \bibnamefont
  {Song}}, \bibinfo {author} {\bibfnamefont {J.-S.}\ \bibnamefont {Jeong}},
  \bibinfo {author} {\bibfnamefont {B.}~\bibnamefont {Pan}}, \bibinfo {author}
  {\bibfnamefont {X.}~\bibnamefont {Zhang}}, \bibinfo {author} {\bibfnamefont
  {J.}~\bibnamefont {Xia}}, \bibinfo {author} {\bibfnamefont {S.}~\bibnamefont
  {Cha}}, \bibinfo {author} {\bibfnamefont {T.-E.}\ \bibnamefont {Park}},
  \bibinfo {author} {\bibfnamefont {K.}~\bibnamefont {Kim}}, \bibinfo {author}
  {\bibfnamefont {S.}~\bibnamefont {Finizio}}, \bibinfo {author} {\bibfnamefont
  {J.}~\bibnamefont {Raabe}}, \bibinfo {author} {\bibfnamefont
  {J.}~\bibnamefont {Chang}}, \bibinfo {author} {\bibfnamefont
  {Y.}~\bibnamefont {Zhou}}, \bibinfo {author} {\bibfnamefont {W.}~\bibnamefont
  {Zhao}}, \bibinfo {author} {\bibfnamefont {W.}~\bibnamefont {Kang}}, \bibinfo
  {author} {\bibfnamefont {H.}~\bibnamefont {Ju}},\ and\ \bibinfo {author}
  {\bibfnamefont {S.}~\bibnamefont {Woo}},\ }\bibfield  {title} {\bibinfo
  {title} {Skyrmion-based artificial synapses for neuromorphic computing},\
  }\href {https://doi.org/10.1038/s41928-020-0385-0} {\bibfield  {journal}
  {\bibinfo  {journal} {Nature Electronics}\ }\textbf {\bibinfo {volume} {3}},\
  \bibinfo {pages} {148} (\bibinfo {year} {2020})}\BibitemShut {NoStop}%
\bibitem [{\citenamefont {Sucksmith}\ and\ \citenamefont
  {Thompson}(1954)}]{Sucksmith1954}%
  \BibitemOpen
  \bibfield  {author} {\bibinfo {author} {\bibfnamefont {W.}~\bibnamefont
  {Sucksmith}}\ and\ \bibinfo {author} {\bibfnamefont {J.~E.}\ \bibnamefont
  {Thompson}},\ }\bibfield  {title} {\bibinfo {title} {The magnetic anisotropy
  of cobalt},\ }\href {https://doi.org/10.1098/rspa.1954.0209} {\bibfield
  {journal} {\bibinfo  {journal} {Proceedings of the Royal Society of London.
  Series A. Mathematical and Physical Sciences}\ }\textbf {\bibinfo {volume}
  {225}},\ \bibinfo {pages} {362} (\bibinfo {year} {1954})}\BibitemShut
  {NoStop}%
\bibitem [{\citenamefont {Chacon}\ \emph {et~al.}(2018)\citenamefont {Chacon},
  \citenamefont {Heinen}, \citenamefont {Halder}, \citenamefont {Bauer},
  \citenamefont {Simeth}, \citenamefont {M{\"u}hlbauer}, \citenamefont
  {Berger}, \citenamefont {Garst}, \citenamefont {Rosch},\ and\ \citenamefont
  {Pfleiderer}}]{Chacon2018}%
  \BibitemOpen
  \bibfield  {author} {\bibinfo {author} {\bibfnamefont {A.}~\bibnamefont
  {Chacon}}, \bibinfo {author} {\bibfnamefont {L.}~\bibnamefont {Heinen}},
  \bibinfo {author} {\bibfnamefont {M.}~\bibnamefont {Halder}}, \bibinfo
  {author} {\bibfnamefont {A.}~\bibnamefont {Bauer}}, \bibinfo {author}
  {\bibfnamefont {W.}~\bibnamefont {Simeth}}, \bibinfo {author} {\bibfnamefont
  {S.}~\bibnamefont {M{\"u}hlbauer}}, \bibinfo {author} {\bibfnamefont
  {H.}~\bibnamefont {Berger}}, \bibinfo {author} {\bibfnamefont
  {M.}~\bibnamefont {Garst}}, \bibinfo {author} {\bibfnamefont
  {A.}~\bibnamefont {Rosch}},\ and\ \bibinfo {author} {\bibfnamefont
  {C.}~\bibnamefont {Pfleiderer}},\ }\bibfield  {title} {\bibinfo {title}
  {Observation of two independent skyrmion phases in a chiral magnetic
  material},\ }\href {https://doi.org/10.1038/s41567-018-0184-y} {\bibfield
  {journal} {\bibinfo  {journal} {Nature Physics}\ }\textbf {\bibinfo {volume}
  {14}},\ \bibinfo {pages} {936} (\bibinfo {year} {2018})}\BibitemShut
  {NoStop}%
\bibitem [{\citenamefont {Bannenberg}\ \emph {et~al.}(2019)\citenamefont
  {Bannenberg}, \citenamefont {Sadykov}, \citenamefont {Dalgliesh},
  \citenamefont {Goodway}, \citenamefont {Schlagel}, \citenamefont {Lograsso},
  \citenamefont {Falus}, \citenamefont {Leli\`evre-Berna}, \citenamefont
  {Leonov},\ and\ \citenamefont {Pappas}}]{Bannenberg2019}%
  \BibitemOpen
  \bibfield  {author} {\bibinfo {author} {\bibfnamefont {L.~J.}\ \bibnamefont
  {Bannenberg}}, \bibinfo {author} {\bibfnamefont {R.}~\bibnamefont {Sadykov}},
  \bibinfo {author} {\bibfnamefont {R.~M.}\ \bibnamefont {Dalgliesh}}, \bibinfo
  {author} {\bibfnamefont {C.}~\bibnamefont {Goodway}}, \bibinfo {author}
  {\bibfnamefont {D.~L.}\ \bibnamefont {Schlagel}}, \bibinfo {author}
  {\bibfnamefont {T.~A.}\ \bibnamefont {Lograsso}}, \bibinfo {author}
  {\bibfnamefont {P.}~\bibnamefont {Falus}}, \bibinfo {author} {\bibfnamefont
  {E.}~\bibnamefont {Leli\`evre-Berna}}, \bibinfo {author} {\bibfnamefont
  {A.~O.}\ \bibnamefont {Leonov}},\ and\ \bibinfo {author} {\bibfnamefont
  {C.}~\bibnamefont {Pappas}},\ }\bibfield  {title} {\bibinfo {title}
  {Skyrmions and spirals in \ce{MnSi} under hydrostatic pressure},\ }\href
  {https://doi.org/10.1103/PhysRevB.100.054447} {\bibfield  {journal} {\bibinfo
   {journal} {Phys. Rev. B}\ }\textbf {\bibinfo {volume} {100}},\ \bibinfo
  {pages} {054447} (\bibinfo {year} {2019})}\BibitemShut {NoStop}%
\bibitem [{\citenamefont {Halder}\ \emph {et~al.}(2018)\citenamefont {Halder},
  \citenamefont {Chacon}, \citenamefont {Bauer}, \citenamefont {Simeth},
  \citenamefont {M\"uhlbauer}, \citenamefont {Berger}, \citenamefont {Heinen},
  \citenamefont {Garst}, \citenamefont {Rosch},\ and\ \citenamefont
  {Pfleiderer}}]{Halder2018}%
  \BibitemOpen
  \bibfield  {author} {\bibinfo {author} {\bibfnamefont {M.}~\bibnamefont
  {Halder}}, \bibinfo {author} {\bibfnamefont {A.}~\bibnamefont {Chacon}},
  \bibinfo {author} {\bibfnamefont {A.}~\bibnamefont {Bauer}}, \bibinfo
  {author} {\bibfnamefont {W.}~\bibnamefont {Simeth}}, \bibinfo {author}
  {\bibfnamefont {S.}~\bibnamefont {M\"uhlbauer}}, \bibinfo {author}
  {\bibfnamefont {H.}~\bibnamefont {Berger}}, \bibinfo {author} {\bibfnamefont
  {L.}~\bibnamefont {Heinen}}, \bibinfo {author} {\bibfnamefont
  {M.}~\bibnamefont {Garst}}, \bibinfo {author} {\bibfnamefont
  {A.}~\bibnamefont {Rosch}},\ and\ \bibinfo {author} {\bibfnamefont
  {C.}~\bibnamefont {Pfleiderer}},\ }\bibfield  {title} {\bibinfo {title}
  {Thermodynamic evidence of a second skyrmion lattice phase and tilted conical
  phase in \ce{Cu2OSeO3}},\ }\href {https://doi.org/10.1103/PhysRevB.98.144429}
  {\bibfield  {journal} {\bibinfo  {journal} {Phys. Rev. B}\ }\textbf {\bibinfo
  {volume} {98}},\ \bibinfo {pages} {144429} (\bibinfo {year}
  {2018})}\BibitemShut {NoStop}%
\bibitem [{\citenamefont {Aqeel}\ \emph {et~al.}(2021)\citenamefont {Aqeel},
  \citenamefont {Sahliger}, \citenamefont {Taniguchi}, \citenamefont {M\"andl},
  \citenamefont {Mettus}, \citenamefont {Berger}, \citenamefont {Bauer},
  \citenamefont {Garst}, \citenamefont {Pfleiderer},\ and\ \citenamefont
  {Back}}]{Aqeel2021}%
  \BibitemOpen
  \bibfield  {author} {\bibinfo {author} {\bibfnamefont {A.}~\bibnamefont
  {Aqeel}}, \bibinfo {author} {\bibfnamefont {J.}~\bibnamefont {Sahliger}},
  \bibinfo {author} {\bibfnamefont {T.}~\bibnamefont {Taniguchi}}, \bibinfo
  {author} {\bibfnamefont {S.}~\bibnamefont {M\"andl}}, \bibinfo {author}
  {\bibfnamefont {D.}~\bibnamefont {Mettus}}, \bibinfo {author} {\bibfnamefont
  {H.}~\bibnamefont {Berger}}, \bibinfo {author} {\bibfnamefont
  {A.}~\bibnamefont {Bauer}}, \bibinfo {author} {\bibfnamefont
  {M.}~\bibnamefont {Garst}}, \bibinfo {author} {\bibfnamefont
  {C.}~\bibnamefont {Pfleiderer}},\ and\ \bibinfo {author} {\bibfnamefont
  {C.~H.}\ \bibnamefont {Back}},\ }\bibfield  {title} {\bibinfo {title}
  {Microwave spectroscopy of the low-temperature skyrmion state in
  \ce{Cu2OSeO3}},\ }\href {https://doi.org/10.1103/PhysRevLett.126.017202}
  {\bibfield  {journal} {\bibinfo  {journal} {Phys. Rev. Lett.}\ }\textbf
  {\bibinfo {volume} {126}},\ \bibinfo {pages} {017202} (\bibinfo {year}
  {2021})}\BibitemShut {NoStop}%
\bibitem [{\citenamefont {Qian}\ \emph {et~al.}(2018)\citenamefont {Qian},
  \citenamefont {Bannenberg}, \citenamefont {Wilhelm}, \citenamefont
  {Chaboussant}, \citenamefont {Debeer-Schmitt}, \citenamefont {Schmidt},
  \citenamefont {Aqeel}, \citenamefont {Palstra}, \citenamefont {Br{\"u}ck},
  \citenamefont {Lefering}, \citenamefont {Pappas}, \citenamefont {Mostovoy},\
  and\ \citenamefont {Leonov}}]{Qian2018}%
  \BibitemOpen
  \bibfield  {author} {\bibinfo {author} {\bibfnamefont {F.}~\bibnamefont
  {Qian}}, \bibinfo {author} {\bibfnamefont {L.~J.}\ \bibnamefont
  {Bannenberg}}, \bibinfo {author} {\bibfnamefont {H.}~\bibnamefont {Wilhelm}},
  \bibinfo {author} {\bibfnamefont {G.}~\bibnamefont {Chaboussant}}, \bibinfo
  {author} {\bibfnamefont {L.~M.}\ \bibnamefont {Debeer-Schmitt}}, \bibinfo
  {author} {\bibfnamefont {M.~P.}\ \bibnamefont {Schmidt}}, \bibinfo {author}
  {\bibfnamefont {A.}~\bibnamefont {Aqeel}}, \bibinfo {author} {\bibfnamefont
  {T.~T.~M.}\ \bibnamefont {Palstra}}, \bibinfo {author} {\bibfnamefont
  {E.}~\bibnamefont {Br{\"u}ck}}, \bibinfo {author} {\bibfnamefont {A.~J.~E.}\
  \bibnamefont {Lefering}}, \bibinfo {author} {\bibfnamefont {C.}~\bibnamefont
  {Pappas}}, \bibinfo {author} {\bibfnamefont {M.}~\bibnamefont {Mostovoy}},\
  and\ \bibinfo {author} {\bibfnamefont {A.~O.}\ \bibnamefont {Leonov}},\
  }\bibfield  {title} {\bibinfo {title} {New magnetic phase of the chiral
  skyrmion material \ce{Cu2OSeO3}},\ }\href@noop {} {\bibfield  {journal}
  {\bibinfo  {journal} {Science Advances}\ }\textbf {\bibinfo {volume} {4}}
  (\bibinfo {year} {2018})}\BibitemShut {NoStop}%
\bibitem [{\citenamefont {Preißinger}\ \emph {et~al.}(2020)\citenamefont
  {Preißinger}, \citenamefont {Karube}, \citenamefont {Ehlers}, \citenamefont
  {Szigeti}, \citenamefont {von Nidda}, \citenamefont {White}, \citenamefont
  {Ukleev}, \citenamefont {Rønnow}, \citenamefont {Tokunaga}, \citenamefont
  {Kikkawa}, \citenamefont {Tokura}, \citenamefont {Taguchi},\ and\
  \citenamefont {Kézsmárki}}]{preissinger2020vital}%
  \BibitemOpen
  \bibfield  {author} {\bibinfo {author} {\bibfnamefont {M.}~\bibnamefont
  {Preißinger}}, \bibinfo {author} {\bibfnamefont {K.}~\bibnamefont {Karube}},
  \bibinfo {author} {\bibfnamefont {D.}~\bibnamefont {Ehlers}}, \bibinfo
  {author} {\bibfnamefont {B.}~\bibnamefont {Szigeti}}, \bibinfo {author}
  {\bibfnamefont {H.~A.~K.}\ \bibnamefont {von Nidda}}, \bibinfo {author}
  {\bibfnamefont {J.~S.}\ \bibnamefont {White}}, \bibinfo {author}
  {\bibfnamefont {V.}~\bibnamefont {Ukleev}}, \bibinfo {author} {\bibfnamefont
  {H.~M.}\ \bibnamefont {Rønnow}}, \bibinfo {author} {\bibfnamefont
  {Y.}~\bibnamefont {Tokunaga}}, \bibinfo {author} {\bibfnamefont
  {A.}~\bibnamefont {Kikkawa}}, \bibinfo {author} {\bibfnamefont
  {Y.}~\bibnamefont {Tokura}}, \bibinfo {author} {\bibfnamefont
  {Y.}~\bibnamefont {Taguchi}},\ and\ \bibinfo {author} {\bibfnamefont
  {I.}~\bibnamefont {Kézsmárki}},\ }\bibfield  {title} {\bibinfo {title}
  {Vital role of anisotropy in cubic chiral skyrmion hosts},\ }\href@noop {} {\
   (\bibinfo {year} {2020})},\ \Eprint {https://arxiv.org/abs/2011.05967}
  {arXiv:2011.05967 [cond-mat.str-el]} \BibitemShut {NoStop}%
\bibitem [{\citenamefont {Birch}\ \emph {et~al.}(2020)\citenamefont {Birch},
  \citenamefont {Moody}, \citenamefont {Wilson}, \citenamefont {Crisanti},
  \citenamefont {Bewley}, \citenamefont {\ifmmode \check{S}\else
  \v{S}\fi{}tefan\ifmmode \check{c}\else \v{c}\fi{}i\ifmmode~\check{c}\else
  \v{c}\fi{}}, \citenamefont {Balakrishnan}, \citenamefont {Fan}, \citenamefont
  {Steadman}, \citenamefont {Alba~Venero}, \citenamefont {Cubitt},\ and\
  \citenamefont {Hatton}}]{Birch2020}%
  \BibitemOpen
  \bibfield  {author} {\bibinfo {author} {\bibfnamefont {M.~T.}\ \bibnamefont
  {Birch}}, \bibinfo {author} {\bibfnamefont {S.~H.}\ \bibnamefont {Moody}},
  \bibinfo {author} {\bibfnamefont {M.~N.}\ \bibnamefont {Wilson}}, \bibinfo
  {author} {\bibfnamefont {M.}~\bibnamefont {Crisanti}}, \bibinfo {author}
  {\bibfnamefont {O.}~\bibnamefont {Bewley}}, \bibinfo {author} {\bibfnamefont
  {A.}~\bibnamefont {\ifmmode \check{S}\else \v{S}\fi{}tefan\ifmmode
  \check{c}\else \v{c}\fi{}i\ifmmode~\check{c}\else \v{c}\fi{}}}, \bibinfo
  {author} {\bibfnamefont {G.}~\bibnamefont {Balakrishnan}}, \bibinfo {author}
  {\bibfnamefont {R.}~\bibnamefont {Fan}}, \bibinfo {author} {\bibfnamefont
  {P.}~\bibnamefont {Steadman}}, \bibinfo {author} {\bibfnamefont
  {D.}~\bibnamefont {Alba~Venero}}, \bibinfo {author} {\bibfnamefont
  {R.}~\bibnamefont {Cubitt}},\ and\ \bibinfo {author} {\bibfnamefont {P.~D.}\
  \bibnamefont {Hatton}},\ }\bibfield  {title} {\bibinfo {title}
  {Anisotropy-induced depinning in the \ce{Zn}-substituted skyrmion host
  \ce{Cu2OSeO3}},\ }\href {https://doi.org/10.1103/PhysRevB.102.104424}
  {\bibfield  {journal} {\bibinfo  {journal} {Phys. Rev. B}\ }\textbf {\bibinfo
  {volume} {102}},\ \bibinfo {pages} {104424} (\bibinfo {year}
  {2020})}\BibitemShut {NoStop}%
\bibitem [{\citenamefont {\ifmmode \check{S}\else \v{S}\fi{}tefan\ifmmode
  \check{c}\else \v{c}\fi{}i\ifmmode~\check{c}\else \v{c}\fi{}}\ \emph
  {et~al.}(2018)\citenamefont {\ifmmode \check{S}\else \v{S}\fi{}tefan\ifmmode
  \check{c}\else \v{c}\fi{}i\ifmmode~\check{c}\else \v{c}\fi{}}, \citenamefont
  {Moody}, \citenamefont {Hicken}, \citenamefont {Birch}, \citenamefont
  {Balakrishnan}, \citenamefont {Barnett}, \citenamefont {Crisanti},
  \citenamefont {Evans}, \citenamefont {Holt}, \citenamefont {Franke},
  \citenamefont {Hatton}, \citenamefont {Huddart}, \citenamefont {Lees},
  \citenamefont {Pratt}, \citenamefont {Tang}, \citenamefont {Wilson},
  \citenamefont {Xiao},\ and\ \citenamefont {Lancaster}}]{Moody2018}%
  \BibitemOpen
  \bibfield  {author} {\bibinfo {author} {\bibfnamefont {A.}~\bibnamefont
  {\ifmmode \check{S}\else \v{S}\fi{}tefan\ifmmode \check{c}\else
  \v{c}\fi{}i\ifmmode~\check{c}\else \v{c}\fi{}}}, \bibinfo {author}
  {\bibfnamefont {S.~H.}\ \bibnamefont {Moody}}, \bibinfo {author}
  {\bibfnamefont {T.~J.}\ \bibnamefont {Hicken}}, \bibinfo {author}
  {\bibfnamefont {M.~T.}\ \bibnamefont {Birch}}, \bibinfo {author}
  {\bibfnamefont {G.}~\bibnamefont {Balakrishnan}}, \bibinfo {author}
  {\bibfnamefont {S.~A.}\ \bibnamefont {Barnett}}, \bibinfo {author}
  {\bibfnamefont {M.}~\bibnamefont {Crisanti}}, \bibinfo {author}
  {\bibfnamefont {J.~S.~O.}\ \bibnamefont {Evans}}, \bibinfo {author}
  {\bibfnamefont {S.~J.~R.}\ \bibnamefont {Holt}}, \bibinfo {author}
  {\bibfnamefont {K.~J.~A.}\ \bibnamefont {Franke}}, \bibinfo {author}
  {\bibfnamefont {P.~D.}\ \bibnamefont {Hatton}}, \bibinfo {author}
  {\bibfnamefont {B.~M.}\ \bibnamefont {Huddart}}, \bibinfo {author}
  {\bibfnamefont {M.~R.}\ \bibnamefont {Lees}}, \bibinfo {author}
  {\bibfnamefont {F.~L.}\ \bibnamefont {Pratt}}, \bibinfo {author}
  {\bibfnamefont {C.~C.}\ \bibnamefont {Tang}}, \bibinfo {author}
  {\bibfnamefont {M.~N.}\ \bibnamefont {Wilson}}, \bibinfo {author}
  {\bibfnamefont {F.}~\bibnamefont {Xiao}},\ and\ \bibinfo {author}
  {\bibfnamefont {T.}~\bibnamefont {Lancaster}},\ }\bibfield  {title} {\bibinfo
  {title} {Origin of skyrmion lattice phase splitting in \ce{Zn}-substituted
  \ce{Cu2OSeO3}},\ }\href {https://doi.org/10.1103/PhysRevMaterials.2.111402}
  {\bibfield  {journal} {\bibinfo  {journal} {Phys. Rev. Materials}\ }\textbf
  {\bibinfo {volume} {2}},\ \bibinfo {pages} {111402} (\bibinfo {year}
  {2018})}\BibitemShut {NoStop}%
\bibitem [{\citenamefont {M\"uhlbauer}\ \emph {et~al.}(2019)\citenamefont
  {M\"uhlbauer}, \citenamefont {Honecker}, \citenamefont {P\'erigo},
  \citenamefont {Bergner}, \citenamefont {Disch}, \citenamefont {Heinemann},
  \citenamefont {Erokhin}, \citenamefont {Berkov}, \citenamefont {Leighton},
  \citenamefont {Eskildsen},\ and\ \citenamefont {Michels}}]{MB2019}%
  \BibitemOpen
  \bibfield  {author} {\bibinfo {author} {\bibfnamefont {S.}~\bibnamefont
  {M\"uhlbauer}}, \bibinfo {author} {\bibfnamefont {D.}~\bibnamefont
  {Honecker}}, \bibinfo {author} {\bibfnamefont {E.~A.}\ \bibnamefont
  {P\'erigo}}, \bibinfo {author} {\bibfnamefont {F.}~\bibnamefont {Bergner}},
  \bibinfo {author} {\bibfnamefont {S.}~\bibnamefont {Disch}}, \bibinfo
  {author} {\bibfnamefont {A.}~\bibnamefont {Heinemann}}, \bibinfo {author}
  {\bibfnamefont {S.}~\bibnamefont {Erokhin}}, \bibinfo {author} {\bibfnamefont
  {D.}~\bibnamefont {Berkov}}, \bibinfo {author} {\bibfnamefont
  {C.}~\bibnamefont {Leighton}}, \bibinfo {author} {\bibfnamefont {M.~R.}\
  \bibnamefont {Eskildsen}},\ and\ \bibinfo {author} {\bibfnamefont
  {A.}~\bibnamefont {Michels}},\ }\bibfield  {title} {\bibinfo {title}
  {Magnetic small-angle neutron scattering},\ }\href
  {https://doi.org/10.1103/RevModPhys.91.015004} {\bibfield  {journal}
  {\bibinfo  {journal} {Rev. Mod. Phys.}\ }\textbf {\bibinfo {volume} {91}},\
  \bibinfo {pages} {015004} (\bibinfo {year} {2019})}\BibitemShut {NoStop}%
\bibitem [{\citenamefont {Arnold}\ \emph {et~al.}(2014)\citenamefont {Arnold},
  \citenamefont {Bilheux}, \citenamefont {Borreguero}, \citenamefont {Buts},
  \citenamefont {Campbell}, \citenamefont {Chapon}, \citenamefont {Doucet},
  \citenamefont {Draper}, \citenamefont {{Ferraz Leal}}, \citenamefont {Gigg},
  \citenamefont {Lynch}, \citenamefont {Markvardsen}, \citenamefont
  {Mikkelson}, \citenamefont {Mikkelson}, \citenamefont {Miller}, \citenamefont
  {Palmen}, \citenamefont {Parker}, \citenamefont {Passos}, \citenamefont
  {Perring}, \citenamefont {Peterson}, \citenamefont {Ren}, \citenamefont
  {Reuter}, \citenamefont {Savici}, \citenamefont {Taylor}, \citenamefont
  {Taylor}, \citenamefont {Tolchenov}, \citenamefont {Zhou},\ and\
  \citenamefont {Zikovsky}}]{ARNOLD2014156}%
  \BibitemOpen
  \bibfield  {author} {\bibinfo {author} {\bibfnamefont {O.}~\bibnamefont
  {Arnold}}, \bibinfo {author} {\bibfnamefont {J.}~\bibnamefont {Bilheux}},
  \bibinfo {author} {\bibfnamefont {J.}~\bibnamefont {Borreguero}}, \bibinfo
  {author} {\bibfnamefont {A.}~\bibnamefont {Buts}}, \bibinfo {author}
  {\bibfnamefont {S.}~\bibnamefont {Campbell}}, \bibinfo {author}
  {\bibfnamefont {L.}~\bibnamefont {Chapon}}, \bibinfo {author} {\bibfnamefont
  {M.}~\bibnamefont {Doucet}}, \bibinfo {author} {\bibfnamefont
  {N.}~\bibnamefont {Draper}}, \bibinfo {author} {\bibfnamefont
  {R.}~\bibnamefont {{Ferraz Leal}}}, \bibinfo {author} {\bibfnamefont
  {M.}~\bibnamefont {Gigg}}, \bibinfo {author} {\bibfnamefont {V.}~\bibnamefont
  {Lynch}}, \bibinfo {author} {\bibfnamefont {A.}~\bibnamefont {Markvardsen}},
  \bibinfo {author} {\bibfnamefont {D.}~\bibnamefont {Mikkelson}}, \bibinfo
  {author} {\bibfnamefont {R.}~\bibnamefont {Mikkelson}}, \bibinfo {author}
  {\bibfnamefont {R.}~\bibnamefont {Miller}}, \bibinfo {author} {\bibfnamefont
  {K.}~\bibnamefont {Palmen}}, \bibinfo {author} {\bibfnamefont
  {P.}~\bibnamefont {Parker}}, \bibinfo {author} {\bibfnamefont
  {G.}~\bibnamefont {Passos}}, \bibinfo {author} {\bibfnamefont
  {T.}~\bibnamefont {Perring}}, \bibinfo {author} {\bibfnamefont
  {P.}~\bibnamefont {Peterson}}, \bibinfo {author} {\bibfnamefont
  {S.}~\bibnamefont {Ren}}, \bibinfo {author} {\bibfnamefont {M.}~\bibnamefont
  {Reuter}}, \bibinfo {author} {\bibfnamefont {A.}~\bibnamefont {Savici}},
  \bibinfo {author} {\bibfnamefont {J.}~\bibnamefont {Taylor}}, \bibinfo
  {author} {\bibfnamefont {R.}~\bibnamefont {Taylor}}, \bibinfo {author}
  {\bibfnamefont {R.}~\bibnamefont {Tolchenov}}, \bibinfo {author}
  {\bibfnamefont {W.}~\bibnamefont {Zhou}},\ and\ \bibinfo {author}
  {\bibfnamefont {J.}~\bibnamefont {Zikovsky}},\ }\bibfield  {title} {\bibinfo
  {title} {Mantid—data analysis and visualization package for neutron
  scattering and $\mu$\ce{SR} experiments},\ }\href
  {https://doi.org/https://doi.org/10.1016/j.nima.2014.07.029} {\bibfield
  {journal} {\bibinfo  {journal} {Nuclear Instruments and Methods in Physics
  Research Section A: Accelerators, Spectrometers, Detectors and Associated
  Equipment}\ }\textbf {\bibinfo {volume} {764}},\ \bibinfo {pages} {156}
  (\bibinfo {year} {2014})}\BibitemShut {NoStop}%
\bibitem [{\citenamefont {Landau}\ and\ \citenamefont
  {Lifshitz}(1980)}]{Landau1980}%
  \BibitemOpen
  \bibfield  {author} {\bibinfo {author} {\bibfnamefont {L.~D.}\ \bibnamefont
  {Landau}}\ and\ \bibinfo {author} {\bibfnamefont {E.~M.}\ \bibnamefont
  {Lifshitz}},\ }\href@noop {} {\emph {\bibinfo {title} {Statistical
  Physics}}},\ \bibinfo {edition} {2nd}\ ed.,\ \bibinfo {series} {1},
  Vol.~\bibinfo {volume} {9}\ (\bibinfo  {publisher} {Pergamon Press},\
  \bibinfo {address} {Pergamon Press Ltd., Headington Hill Hall, Oxford OX3
  0BW, England},\ \bibinfo {year} {1980})\BibitemShut {NoStop}%
\bibitem [{\citenamefont {White}(1983)}]{White2007}%
  \BibitemOpen
  \bibfield  {author} {\bibinfo {author} {\bibfnamefont {R.~M.}\ \bibnamefont
  {White}},\ }\href@noop {} {\emph {\bibinfo {title} {Quantum Theory of
  Magnetism}}}\ (\bibinfo  {publisher} {Springer},\ \bibinfo {address}
  {Berlin},\ \bibinfo {year} {1983})\BibitemShut {NoStop}%
\bibitem [{\citenamefont {Izyumov}(1984)}]{Izyumov_1984}%
  \BibitemOpen
  \bibfield  {author} {\bibinfo {author} {\bibfnamefont {Y.~A.}\ \bibnamefont
  {Izyumov}},\ }\bibfield  {title} {\bibinfo {title} {Modulated, or
  long-periodic, magnetic structures of crystals},\ }\href
  {https://doi.org/10.1070/pu1984v027n11abeh004120} {\bibfield  {journal}
  {\bibinfo  {journal} {Soviet Physics Uspekhi}\ }\textbf {\bibinfo {volume}
  {27}},\ \bibinfo {pages} {845} (\bibinfo {year} {1984})}\BibitemShut
  {NoStop}%
\end{thebibliography}%

\end{document}